\documentclass[twocolumn]{aastex62}
\usepackage{graphicx,natbib}

\received{?, 2018}
\revised{?, 2018}
\accepted{?, 2018}
\submitjournal{ApJ}

\shorttitle{Stellar surface magneto-convection  as a source of astrophysical noise}
\shortauthors{Cegla et al.}
\begin{document}
\title{Stellar Surface Magneto-Convection as a Source of Astrophysical Noise\\ II. Center-to-Limb Parameterisation of  Absorption Line Profiles and Comparison to Observations}

\correspondingauthor{H.~M. Cegla}
\email{h.cegla@unige.ch}

\author{H.~M. Cegla}
\affiliation{Observatoire de Gen\`eve, Universit\`e de Gen\`eve, 51 Chemin des Maillettes, 1290, Versoix, Switzerland}
\affiliation{CHEOPS Fellow, SNSF NCCR-PlanetS}
\affiliation{Astrophysics Research Centre, School of Mathematics \& Physics, Queen's University Belfast, University Road, Belfast BT7 1NN, UK}

\author{C.~A. Watson}
\affiliation{Astrophysics Research Centre, School of Mathematics \& Physics, Queen's University Belfast, University Road, Belfast BT7 1NN, UK}
\author{S. Shelyag}
\affiliation{Department of Mathematics, Physics and Electrical Engineering, Northumbria University, Newcastle upon Tyne, NE1 8ST, UK}
\author{W.~J. Chaplin}
\affiliation{School of Physics and Astronomy, University of Birmingham, Edgbaston, Birmingham B15 2TT, UK}
\affiliation{Stellar Astrophysics Centre, Department of Physics and Astronomy, Aarhus University, Ny Munkegade 120, 8000 Aarhus C, Denmark}
\author{G.~R. Davies}
\affiliation{School of Physics and Astronomy, University of Birmingham, Edgbaston, Birmingham B15 2TT, UK}
\affiliation{Stellar Astrophysics Centre, Department of Physics and Astronomy, Aarhus University, Ny Munkegade 120, 8000 Aarhus C, Denmark}
\author{M. Mathioudakis}
\affiliation{Astrophysics Research Centre, School of Mathematics \& Physics, Queen's University Belfast, University Road, Belfast BT7 1NN, UK}
\author{M.~L. III Palumbo}
\affiliation{Department of Physics and Astronomy, University of North Carolina at Chapel Hill, 120 E Cameron Ave, Chapel Hill, NC 27514, USA}
\author{S.~H. Saar}
\affiliation{Harvard-Smithsonian Center for Astrophysics, 60 Garden Street, Cambridge, MA 02138, USA}
\author{R.~D. Haywood}
\affiliation{Harvard-Smithsonian Center for Astrophysics, 60 Garden Street, Cambridge, MA 02138, USA}
\affiliation{NASA Sagan Fellow}

\begin{abstract}
Manifestations of stellar activity (such as star-spots, plage/faculae, and convective flows) are well known to induce spectroscopic signals often referred to as astrophysical noise by exoplanet hunters. For example, setting an ultimate goal of detecting true Earth-analogs demands reaching radial velocity (RV) precisions of $\sim$9 cm s$^{-1}$. While this is becoming technically feasible with the latest generation of highly stabilised spectrographs, it is astrophysical noise that sets the true fundamental barrier on attainable RV precisions. In this paper we parameterise the impact of solar surface magneto-convection on absorption line profiles, and extend the analysis from the solar disc centre (Paper~I) to the solar limb. Off disc-centre, the plasma flows orthogonal to the granule tops begin to lie along the line-of-sight and those parallel to the granule tops are no longer completely aligned with the observer. Moreover, the granulation is corrugated and the granules can block other granules, as well as the intergranular lane components. Overall, the visible plasma flows and geometry of the corrugated surface significantly impact the resultant line profiles and induce centre-to-limb variations in shape and net position. We detail these herein, and compare to various solar observations. We find our granulation parameterisation can recreate realistic line profiles and induced radial velocity shifts, across the stellar disc, indicative of both those found in computationally heavy radiative 3D magnetohydrodynamical simulations and empirical solar observations. 
\end{abstract}


\section{Introduction}
\label{sec:intro}
There are many phenomena on the surfaces of potentially planet-hosting low-mass, main-sequence stars that prevent them from being perfect, homogeneous spheres. For example, such stars experience stellar surface magneto-convection (granulation), oscillations, flares, dark starspots and bright networks of faculae/plage with regions of (magnetically) suppressed convection. In addition, they also exhibit magnetic activity cycles that drive quasi-periodic variations in the properties of such features, such as their distributions, sizes, filling factors, and lifetimes. From a spectroscopic point of view, these phenomena can alter the observed stellar line profiles, and in turn may be mistakenly interpreted as whole-sale Doppler shifts – referred to as astrophysical noise or stellar `jitter'. The level of astrophysical noise on a typical exoplanet host may range from 10s of cm~s$^{-1}$ to 100s of m~s$^{-1}$ \citep{saar97,schrijver00}; as such, these spurious signals from the star completely swamp the 9 cm~s$^{-1}$ Doppler-reflex motion induced by a true Earth-analogue (and can even mimic some planets, e.g. see \citealt{robertson15}). This is particularly troubling, as our best hope for the confirmation of an Earth-twin is from the spectroscopic follow-up of small planet candidates found in photometric surveys (e.g. Kepler/K2, TESS, and PLATO). Moreover, as spectrographs capable of 10 cm~s$^{-1}$ precision start to come online (e.g. ESPRESSO has recently seen first light\footnote{https://www.eso.org/public/news/eso1739/}), the impact of astrophysical noise will increasingly represent the fundamental limit to the confirmation and characterisation of exoplanets -- and it is this aspect, rather than instrumental capabilities, that will be critical in the search and confirmation of Earth-analogues.

Even `quiet' non-active stars exhibit oscillations and stellar surface magneto-convection/granulation. In particular, pressure (p-) modes give rise to periodic perturbations of the stars, and hence induce Doppler shifts, on dominant timescales of the order of several minutes for Sun-like stars \citep[][and references therein]{christensen-dalsgaard02}. Granulation induces radial velocity (RV) shifts by creating asymmetries in the line profiles from the combination of hot up-rising, blueshifted granules and cool, redshifted, sinking intergranular lanes. This can result in a net convective blueshift since the granules are brighter and cover more surface area \citep[][and references therein]{stein12}. The timescales for granulation are similar to the p-modes, since it is the turbulent motions in the convective envelope that drive the acoustic oscillations, and both are tied to the stellar surface gravity \cite[][and references therein]{kallinger14}. However, since the granules tend to appear and disappear in the same locations, long exposure times do not completely average out the granulation impact. The current mitigation technique for this noise is to simply `beat it down' by adjusting the number of observations and exposure times \citep{dumusque11a}; however, this is extremely costly and may reach a fundamental noise floor. Moreover, oscillations may be easier to `beat down' than granulation, despite their similar time-scales, since they produce narrow-band signals (i.e. higher quality factor, with lower relative damping) that can readily be filtered, whilst the granulation has a pink-noise-like signature spread in frequency. For example, \cite{chaplin18} demonstrate that solar oscillations can be averaged out to $\lesssim$~10~cm~s$^{-1}$ with an exposure time of $\sim$5.4 minutes, while \cite{meunier15} argue that it would take more than one night to average the granulation signal to the 10~cm~s$^{-1}$ level. Consequently, in this series of papers we explore the impact of stellar surface magneto-convection as a source of astrophysical noise. 

In Paper~I \citep{cegla13}, we outlined our technique for characterising photospheric magneto-convection at disc centre. The backbone of this characterisation is a state-of-the-art 3D magnetohydrodynamic (MHD) solar simulation, coupled with detailed wavelength-dependent radiative transfer. Due to the time-intensive nature of detailed radiative diagnostics, producing enough realistic granulation patterns to cover an entire stellar disc with this method is not feasible. As a result, we developed a multi-component parameterisation of granulation at disc centre from a short time-series of simulations, and have shown that we can reconstruct the line profile asymmetries and RV shifts due to the photospheric convective motions found in the MHD-based simulations. The parameterisation is composed of absorption line profiles calculated for granules, magnetic intergranular lanes, non-magnetic intergranular lanes, and magnetic bright points at disc centre. These components were constructed by averaging Fe~I~$6302~\mathrm{\AA}$ magnetically sensitive absorption line profiles output from detailed radiative transport calculations of the solar photosphere. 

In this paper, we extend the multi-component parameterisation of stellar surface magneto-convection from disc centre (Paper~I) towards the stellar limb. In a forthcoming publication (Paper~III), this parameterisation will be used to create new absorption line profiles that represent realistic granulation patterns and create Sun-as-a-star model observations; these artificially noisy model stars will then be used to test astrophysical noise reduction techniques. For a comprehensive overview of general granulation properties, such as granule areas, intensities, lifetimes, and flow characteristics etc., we direct the reader to the HD and MHD simulations of \cite{beeck13a, beeck13, beeck15a, beeck15b} for FGKM stars. Here we specifically focus on parameterising solar simulations, with an average magnetic field of 200~G, in order to understand the centre-to-limb behaviour and enable the future generation of realistic granulation line profiles. 

In Section~\ref{sec:2_80_full_prfls}, we show how the average granulation line profiles from the simulation snapshots change as a function of centre-to-limb angle, and in Section~\ref{sec:2_80_prfls} we break this down into how each of the four granulation components change. We explore the accuracy of the parameterisation across the stellar disc in Section~\ref{sec:2_80_recon} and compare to observations in Section~\ref{subsec:obs}. Finally, we conclude in Section~\ref{sec:conc}. It is important to note that the underlying MHD simulation and radiative transport code have undergone minor changes since Paper~I (see Section~\ref{sec:2_80_full_prfls} for details). 

\section{Centre-to-limb Variations in the Average Absorption Line Profiles}
\label{sec:2_80_full_prfls}
\subsection{Simulation Details}
\label{subsec:sim}
Similar to Paper~I, we use the MURaM code to produce 3D MHD solar surface simulations \citep{vogler05}. The numerical domain has a physical size of 12 x 12 Mm$^{2}$ in the horizontal direction and 2.0 Mm in the vertical direction and is resolved by 480 x 480 x 200 grid cells. Note the simulation box has increased in depth since Paper~I to circumvent any potential issues with boundary conditions, and the equation of state has also been upgraded to OPAL \citep{rogersOPAL, rogersnayfonovOPAL}. The simulation box is inclined in 2$^{\rm{o}}$ steps from disc centre to 80$^{\rm{o}}$. Inclinations beyond 80$^{\rm{o}}$ were computationally unfeasible due to the $1 / \cos(i)$ dependency, requiring a significantly longer ray-path. Nonetheless, we note this corresponds to $\sim$0.91 times the solar radius, and as the small annulus beyond this is heavily limb darkened it contributes less than 5\% to the total integrated flux.
\begin{center}
\begin{figure*}[t]
\begin{center}
\includegraphics[width=8.5cm]{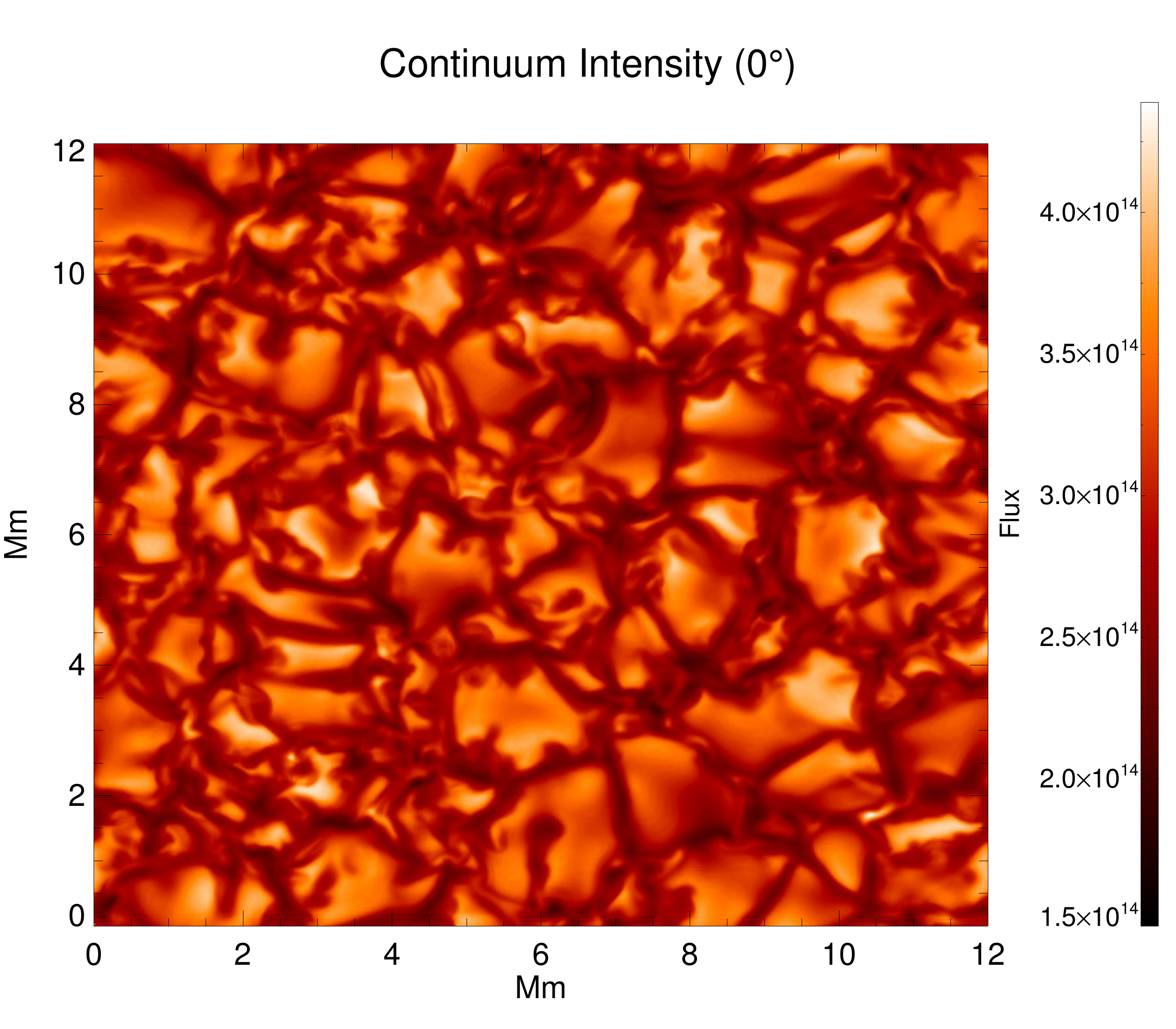}
\includegraphics[width=8.5cm]{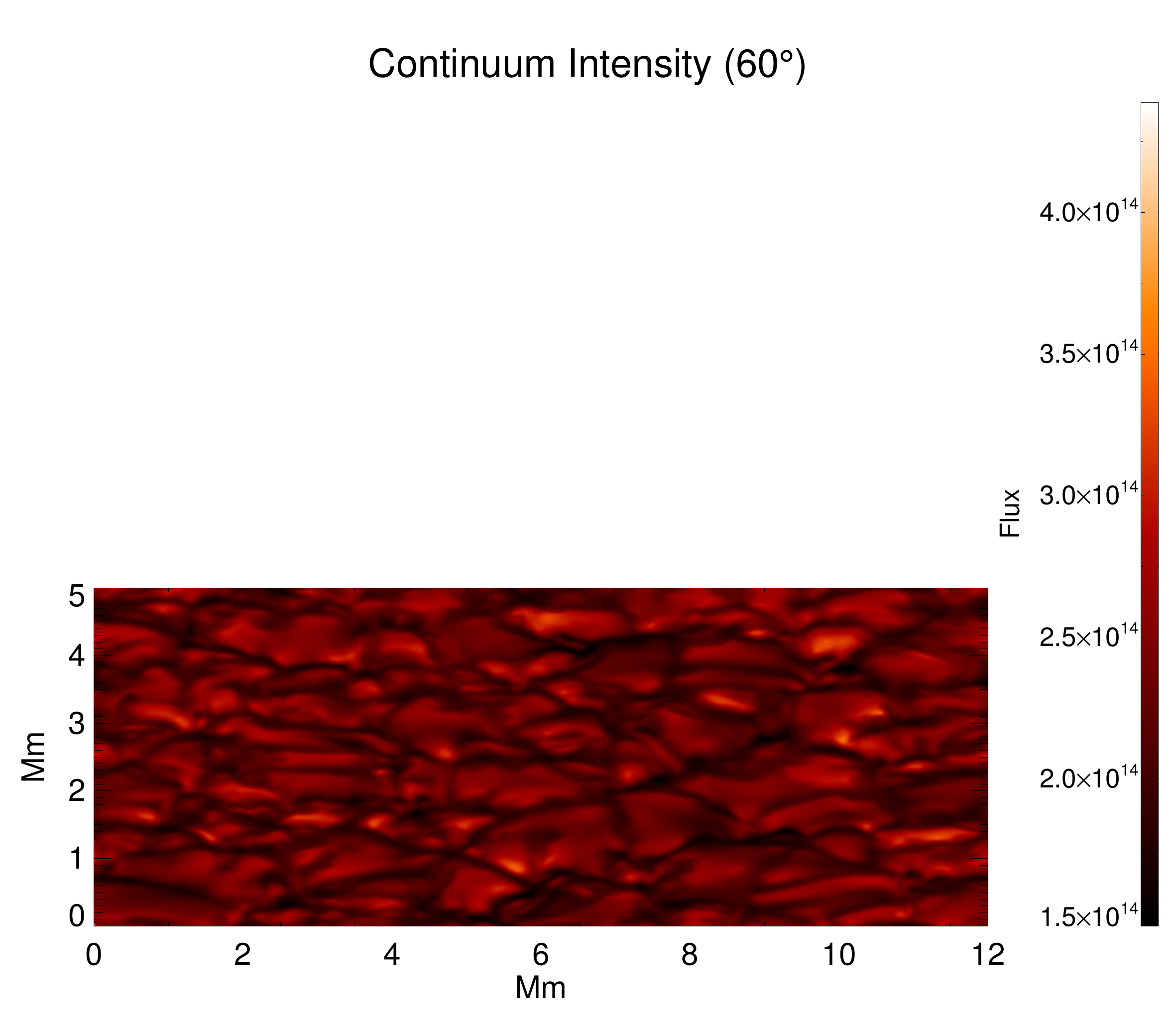}
\includegraphics[width=8.5cm]{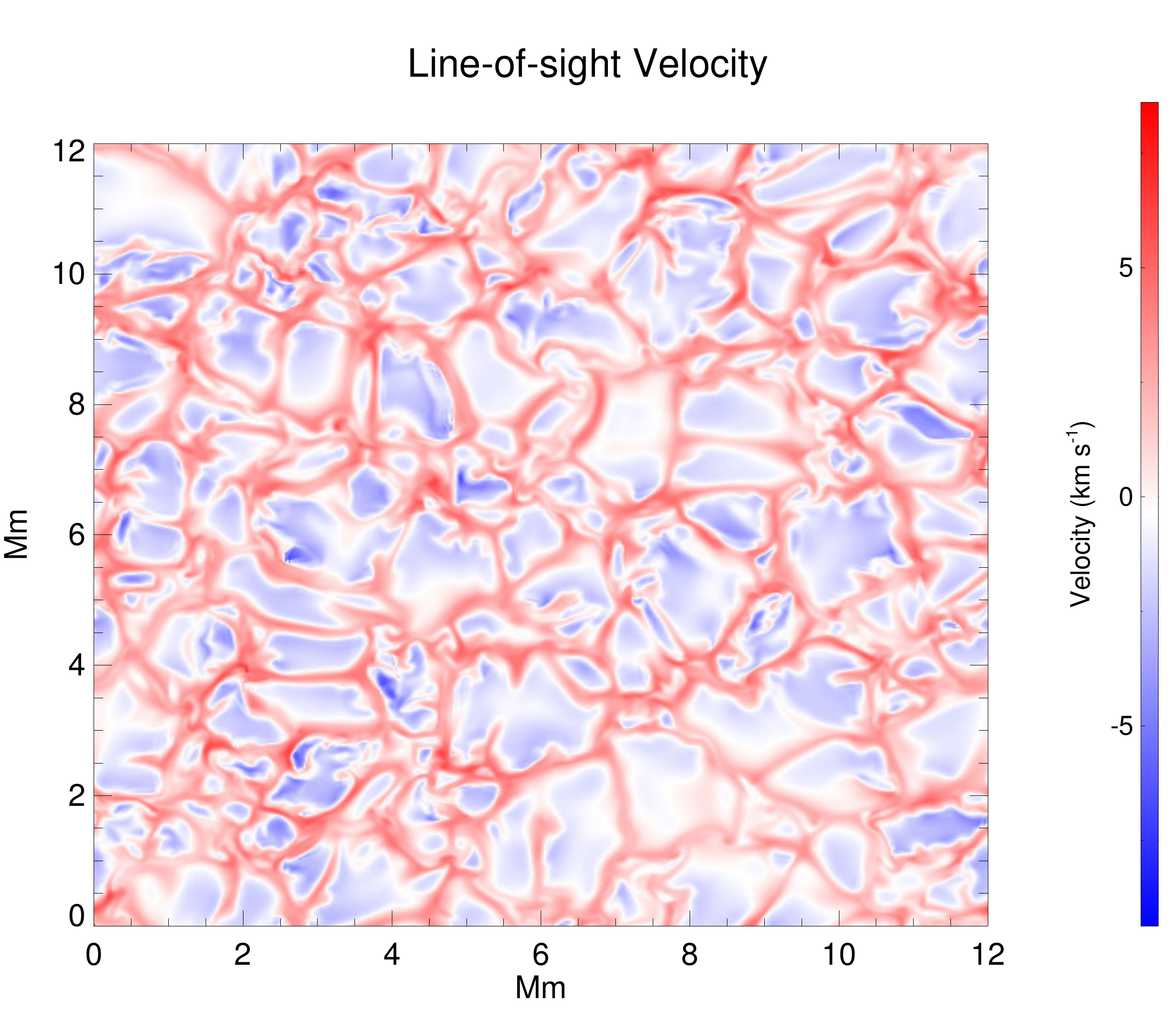}
\includegraphics[width=8.5cm]{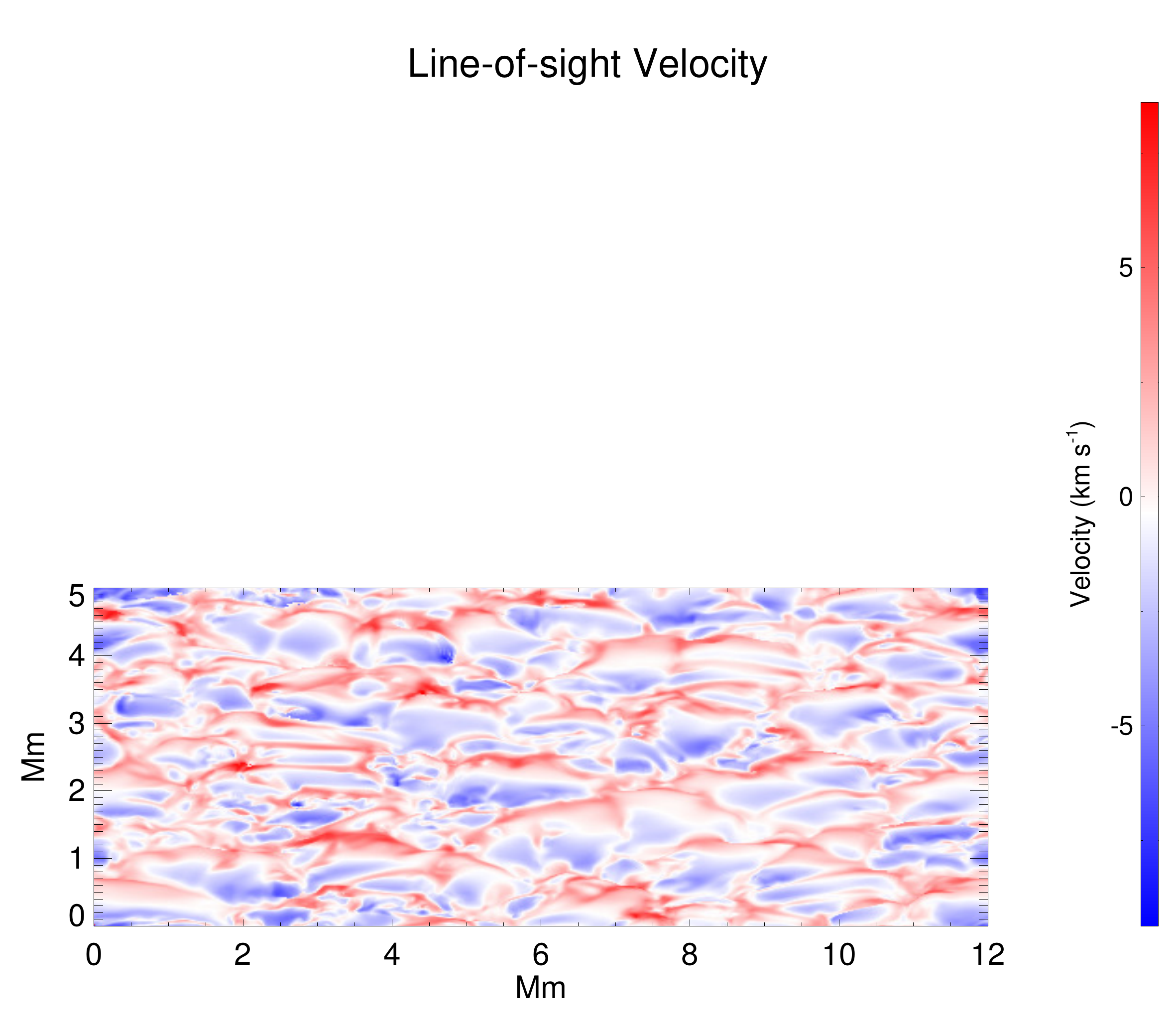}
\caption{Continuum intensities (top) and line-of-sight velocities (bottom) for one simulation snapshot at two different stellar disc centre-to-limb angles: 0$^{\rm{o}}$ (left) and 60$^{\rm{o}}$ (right). Negative and positive velocities denote blueshifts and redshifts, respectively.} 
\label{fig:2_80_cont_vel}
\end{center}
\end{figure*}
\end{center}
\vspace{-10pt}

Again, we have introduced a uniform vertical $200~\mathrm{G}$ magnetic field, and this time we produced a sequence of 201 snapshots. Note that the MHD simulation did not have exactly even time steps, but we have tried to select snapshots with a cadence as close to 30~s as possible, but nearly a third of the snapshots (that are randomly distributed in the series) have a sampling closer to 15~s. This sequence covers approximately 100 minutes of physical time or $\sim$10 - 25 granular lifetimes. See Paper~I for more details. We have used the 1D radiative transport code NICOLE \citep{NICOLE1, NICOLE2}, in conjunction with the MHD simulation, to synthesise the Fe~I~$6302~\mathrm{\AA}$ magnetically sensitive absorption line in local thermodynamic equilibrium (LTE). Similar to Paper~I, we focus only on the Stokes~$I$ component, and resolve a $\pm~0.3~\mathrm{\AA}$ region with 400 points. Note, with NICOLE we have used updated opacities relative to Paper~I. The changes in the MURaM code, and to NICOLE for the line synthesis, were motivated by comparisons with observations (see Sec.~\ref{subsec:obs} for more details). The technical details on the absorption line profile calculations for inclined models and simulated spectro-polarimetry off the solar disc centre are provided in \citet{shelyag14, shelyag15}.

\subsection{Centre-to-limb Variations}
\label{subsec:full_CLV}
Towards the stellar limb the absorption line profiles are subject both to limb darkening and the fact that the granulation makes the surface appear corrugated; this is illustrated in Figure~\ref{fig:2_80_cont_vel} where the continuum intensity and line-of-sight (LOS) velocities of one snapshot are shown for disc centre and near the limb (0$^{\rm{o}}$ and 60$^{\rm{o}}$) -- note blueshifts are denoted by negative velocities and redshifts by positive velocities. An apt analogy for the corrugated surface is to think of the granules as ‘hills’ and the intergranular lanes as ‘valleys’ \citep{dravins08}. As the simulation is inclined, some of the granular `hills' obstruct the intergranular lane `valleys', the sides of the granular walls become visible, and parts of the granule tops can be obscured. Moreover, towards the stellar limb, plasma flows that were perpendicular to the LOS at disc centre are no longer completely orthogonal to the viewer and thus have non-zero RVs. Additionally, the purely vertical flows at disc centre are now inclined with respect to the LOS, and the observed RVs decrease in magnitude as a function of the projected area.    
\begin{center}
\begin{figure}[t!]
\centering
\includegraphics[trim=0.5cm 0.25cm 0.25cm 0.5cm, clip, width=8.5cm]{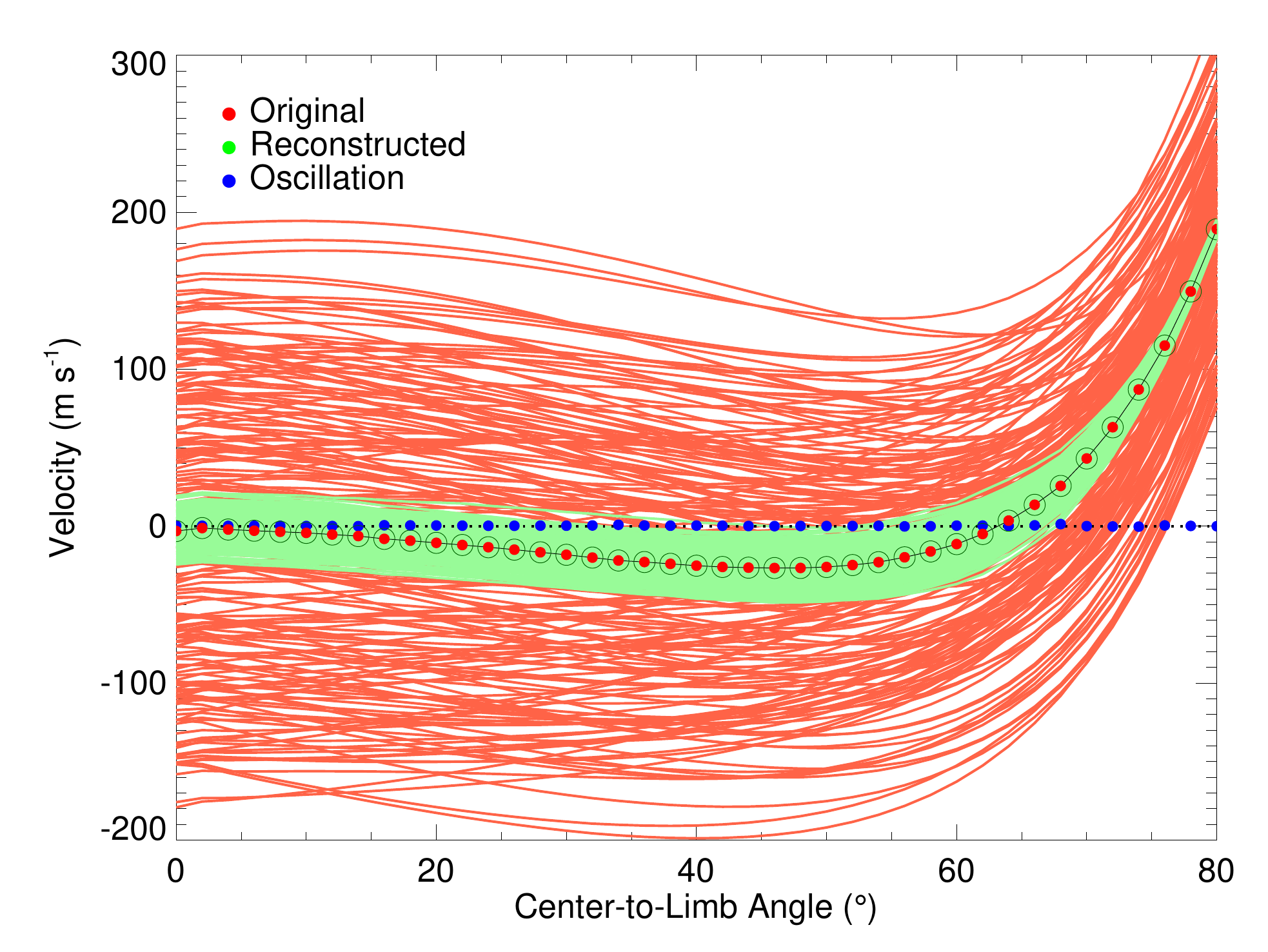}
\caption{Net RVs relative to disc centre for both individual snapshots (lines) and averages over the time-series (points); original profile results are in red, connected by a solid black line, and the reconstructed (granulation-only) profiles are in green (see Sec.~\ref{sec:2_80_prfls} and \ref{sec:2_80_recon} for details). Scatter from individual profiles is due to p-modes and granular evolution (reconstruction only includes granular effects). Also plotted are the oscillation RVs averaged over the entire series (blue points), and a black dotted line at 0 $\mathrm{m~s^{-1}}$~to guide the eye.} 
\label{fig:2_80_rv_0dgr}
\end{figure}
\end{center}
\vspace{-30pt}

\begin{center}
\begin{figure*}[t]
\centering
\includegraphics[trim=1.9cm 0.2cm 1.1cm 1.3cm, clip, scale=0.5]
{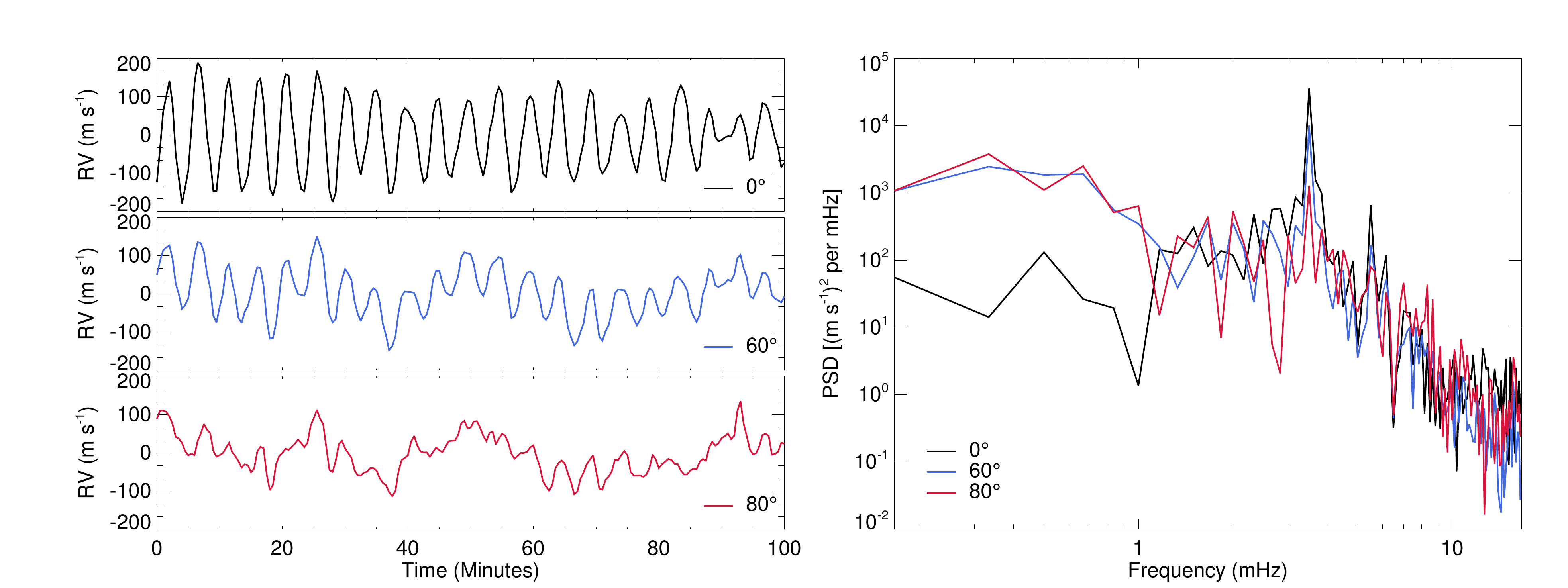}
\caption{Original RVs from the average line profiles from each snapshot over the time-series (left) and their corresponding power spectra (right) for limb angles 0$^{\rm{o}}$, 60$^{\rm{o}}$, 80$^{\rm{o}}$;  these include contributions from both p-mode oscillations and granulation.} 
\label{fig:orgRV}
\end{figure*}
\end{center}
\vspace{-20pt}

Toward the limb, only the near-edge of the granules have high blueshifts as this is where the plasma is starting to flow into the intergranular lane and hence lies more inline with our LOS. The remainder of the granule tops point away from the observer, and plasma flows that were once seen as coming towards the observer now begin to point away, resulting in a decrease in blueshift of the granules as whole. The opposite effect is seen in the intergranular lanes, where some of the downward flowing material is now flowing toward the observer as the tiles are inclined. It is also possible to see the intergranular lanes underneath some of the smaller granules. See Section~\ref{subsec:2_80_CLV} for more details on the responses of individual granulation components. As a consequence of this corrugation, we see differences in line shape, line centre, and filling factor for each of the granulation components as a function of limb angle (discussed in the Section~\ref{sec:2_80_prfls}); this directly results in the average granulation RV also changing as a function of limb angle.

Of particular significance is the centre-to-limb variation (CLV) of the net convective velocities. From solar observations, we can expect a CLV on the 100~m~s$^{-1}$ level, becoming more redshifted toward the stellar limb \citep{dravins82}; this is because plasma flows moving away from the observer are more often seen in front of the hotter (i.e. brighter) plasma above the intergranular lanes, and the flows towards the observer are increasingly blocked from view by granules located in the forefront of the LOS \citep[see][and references therein]{balthasar85,asplund00}. To investigate this relative variation in our simulations, we use the approach outlined in Paper~I to compute the RVs: the average line profile from each individual snapshot in the time-series is cross-correlated against an arbitrarily chosen template profile. The template was created from the parameterised reconstruction described in Section~\ref{sec:2_80_recon}, but merely serves as a reference point. The same disc centre template is used for all limb angles. As such, these net RVs are relative to disc centre and are not absolute. The results from individual snapshots are shown as red lines, and the average over the time-series is shown as red points in Figure~\ref{fig:2_80_rv_0dgr}. The very large variation amongst individual snapshots originates from the RV shifts introduced from p-mode oscillations that are naturally excited by the convection in the simulation box, which are approximately a factor of ten larger in amplitude than the granulation induced shifts (shown in green  -- derived from the granulation parameterisation in Section~\ref{sec:2_80_prfls} and the subsequent line profile reconstruction in Section~\ref{sec:2_80_recon}). As discussed in Section~\ref{subsec:rmv_osc}, these oscillations can largely be averaged out over the time-series, and hence appear near 0 in Figure~\ref{fig:2_80_rv_0dgr} (blue dots); in empirical observations the p-modes may also be averaged to a root-mean-square $\lesssim$~10~cm~s$^{-1}$ with an exposure time of 5.4 minutes for the Sun \citep{chaplin18}. Consequently, we focus on the nature of the time-averaged net RV variation over the stellar disc. 

From our simulations, we find a net redshift near the limb of $\sim$200~m~s$^{-1}$, relative to disc centre -- inline with what we expect from solar observations. Part of this net RV CLV is due to projected area effects. However, the change in net RVs is not only larger in magnitude than the $\cos (i)$ term from the projected area, but also deviates from this effect since the stellar surface is corrugated. From Figure~\ref{fig:2_80_rv_0dgr}, we see an initial increase in the net blueshift, which is likely due to increased contributions from velocity flows on the granule tops that have a LOS component away from disc centre. This increase in blueshift continues until a limb angle of 46$^{\rm{o}}$ ($\mu \approx 0.7$, where $\mu = \cos (\theta)$). Then near the limb ($>60^{\rm{o}}$), we see a rapid increase in the relative redshift, owing to the aforementioned velocity flows seen in the hot/bright plasma above the intergranular lanes. The net impact seen here is further broken down into the impact from individual granulation components in Section~\ref{sec:2_80_prfls}.

\subsection{Oscillations}
\label{subsec:osc_fullprfl}
The magneto-convection in the simulation box naturally excites p-modes and f-modes (surface gravity flows), as happens on the real Sun. We note that while f-modes have been detected at low frequencies in high-resolution data on the Sun, they have very low amplitudes and cannot be detected in Sun-as-a-star data. The various modes both constructively and destructively interfere with one another to induce oscillatory variations in the line profiles. Since the oscillations are not only stochastically excited but are also intrinsically damped by the convection, the signal de-phases on a timescale commensurate with the damping time for each mode. The hypothesis from Paper~I was that the oscillations only induced whole-sale Doppler shifts and did not significantly change the line profile shape \citep{gray05}, and the vertical scatter within the individual line profiles was naively attributed solely to granulation. However, further inspection of the power spectra densities (PSDs), calculated via fast Fourier transforms (FFTs), show clear periodicities in both the RVs and line shape parameters (e.g. line depth, width, and equivalent width etc.) in the classic `5-minute solar oscillation' range ($\sim$3~mHz). Moreover, the p-mode induced\footnote{Identified by their peak periodicities.} Doppler shifts and line shape variations are up to an order of magnitude larger than those induced from convection in our simulation box (e.g. compare RVs in Figures~\ref{fig:orgRV} and \ref{fig:2_80_residrv_time}). Note that a decrease in numerical plasma viscosity in the MHD simulation from Paper~I means the oscillation never completely dampens out in this time-series. 
\begin{figure*}[t]
\centering
\includegraphics[trim={0 4cm 0 5cm}, scale=0.4]{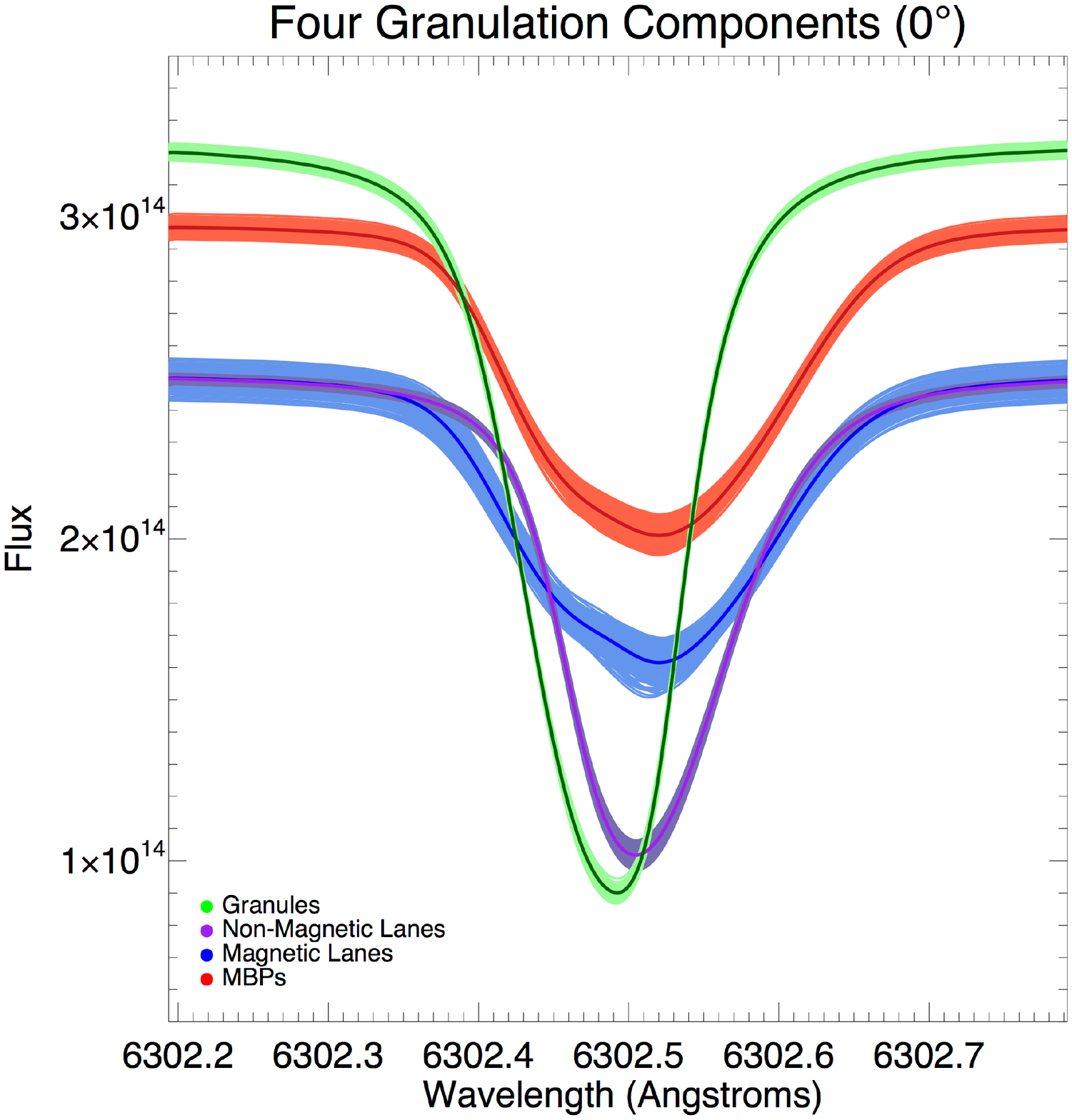}
\includegraphics[trim={0 4cm 0 5cm}, scale=0.4]{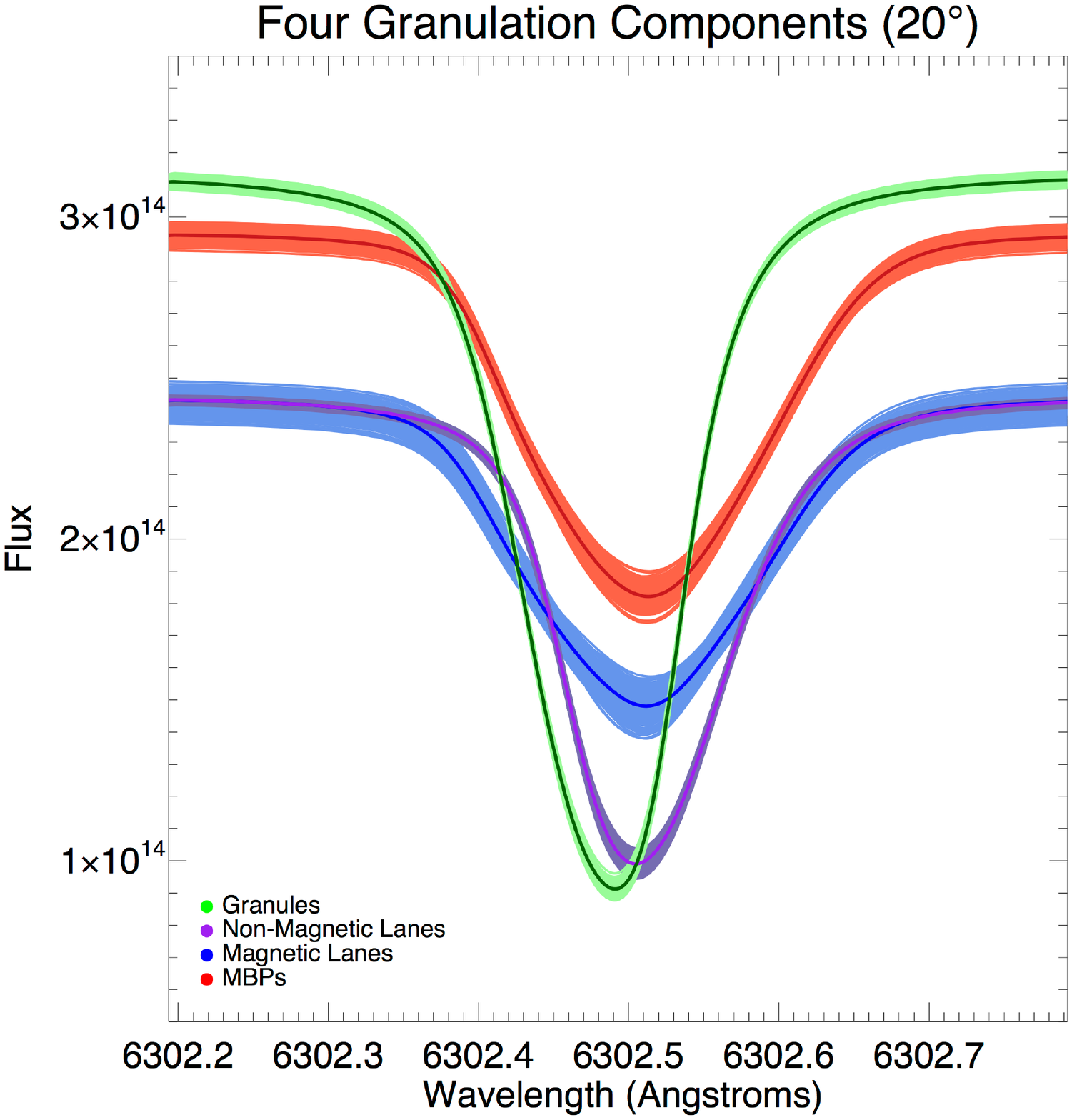}
\includegraphics[trim={0 5cm 0 5cm}, scale=0.4]{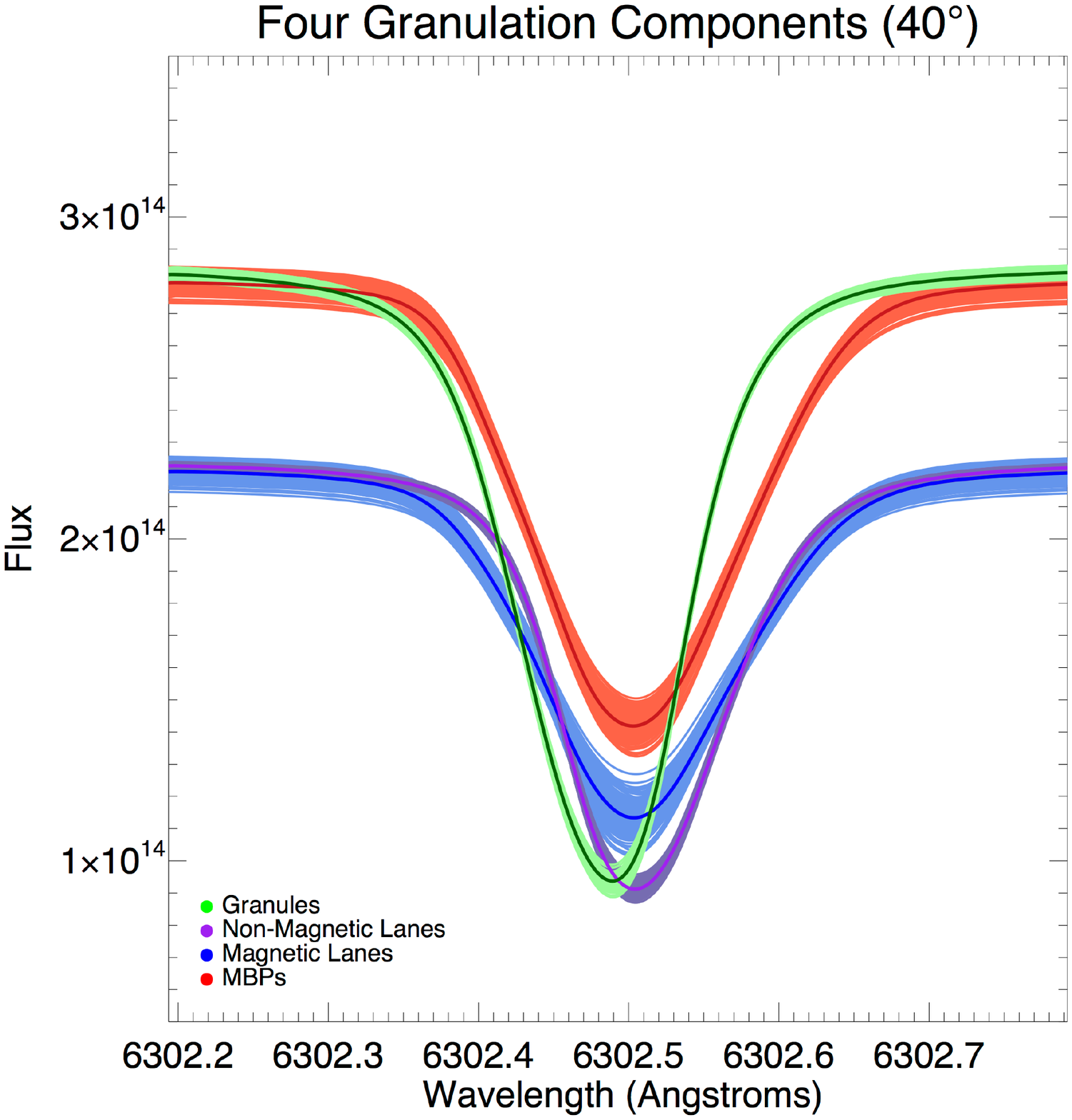}
\includegraphics[trim={0 5cm 0 5cm}, scale=0.4]{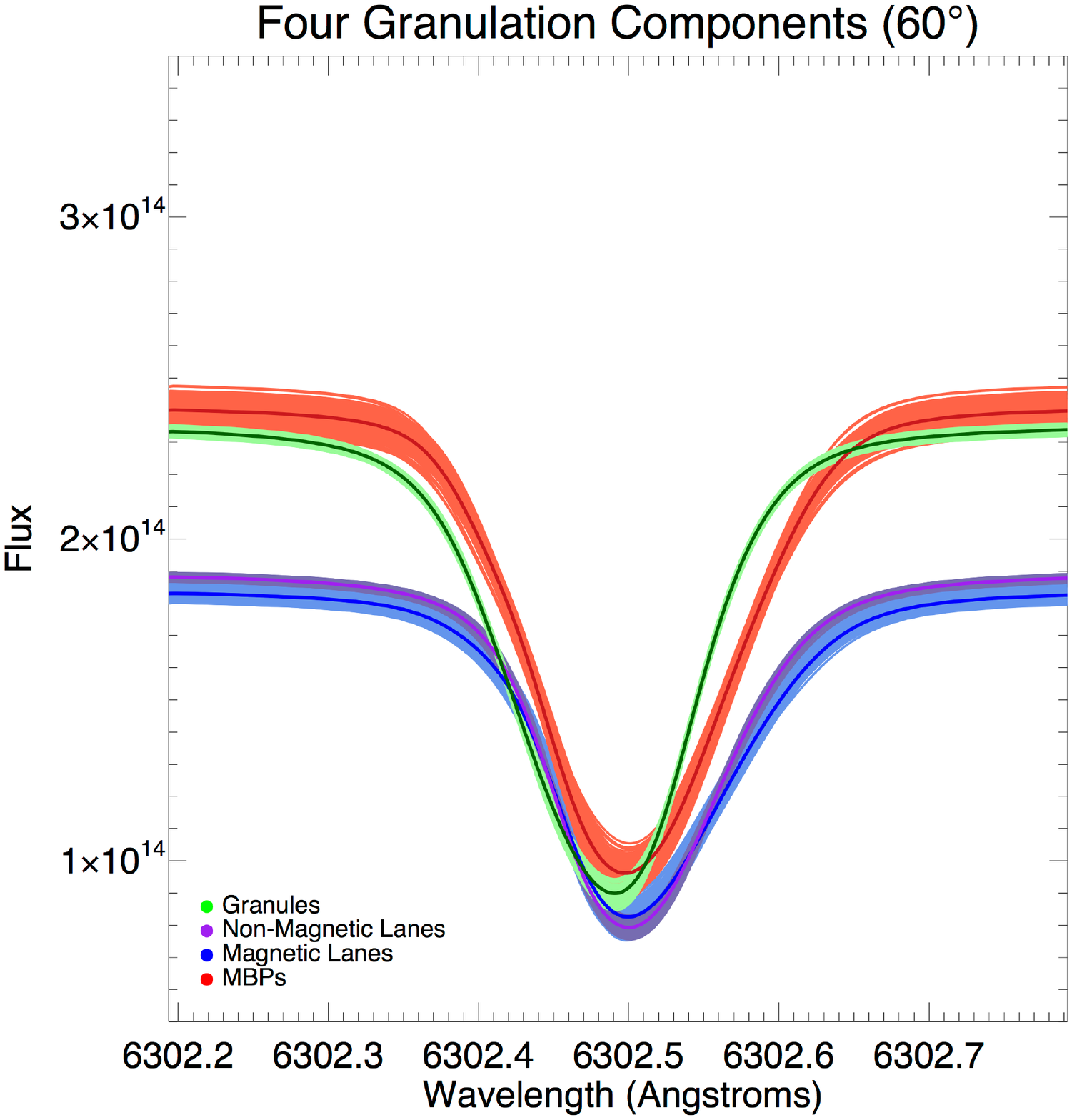}
\caption{Average (thick) and individual (thin) line profiles from the four different contributions to granulation used in the parameterisation: granules (green), MBPs (red), magnetic (blue) and non-magnetic (purple) intergranular lanes, for four different stellar disc centre-to-limb angles. Flux is measured in units of erg s$^{-1}$ cm$^{-2}$ sr$^{-1}$ \AA$^{-1}$.} 
\label{fig:2_80_4comp}
\end{figure*}

From solar observations we expect the amplitude of the p-modes to decrease in amplitude toward the stellar limb \citep{schmidt99}. This is due to two contributing factors: near the limb we see a combination of vertical and horizontal velocity flows, and the inclined LOS means we are viewing areas that are physically higher in the solar atmosphere. In Figure~\ref{fig:orgRV}, we show both the decrease in amplitude of the p-mode oscillation as a function of limb angle, and the decrease in power as seen in the PSD. At 80$^{\rm{o}}$ ($\mu \approx 0.17$), the p-mode amplitudes are barely visible and the power is decreased by nearly 90\% compared to disc centre. These findings are in agreement with \cite{lohner18} and their recent high precision solar observations with the LARS spectrograph. 

\section{The Four Component Model: \\ Across the Stellar Disc}
\label{sec:2_80_prfls}
Here we extend the multi-component parameterisation of solar surface magneto-convection from disc centre in Paper~I towards the stellar limb, in steps of 2$^{\rm{o}}$ up to an angle of 80$^{\rm{o}}$ ($\mu \approx 0.17$). The individual line profiles that correspond to each pixel are still separated into four different granulation components using continuum intensity and magnetic field cuts; we use the same cuts as Paper~I: 0.9 times the average continuum intensity (at each limb angle) of a randomly chosen snapshot and 1~kG times cosine of the limb angle (note that although the average magnetic field of the simulation is $200~\mathrm{G}$, once this is advected into the intergranular lanes by the granules, regions within a given snapshot can easily reach values $>$ 1~kG). The magnetic field cut away from disc centre is not perfect as it assumes that the fields are purely radial from the centre of the star, which is not strictly true; however, the contribution from non-radial magnetic flux is comparatively small so this cut should be reasonable to separate magnetic and non-magnetic components. All snapshots are scaled to the projected area at each limb angle; for the magnetic field we operate on the LOS along a region of constant temperature corresponding to the stellar surface (5800~K). As such, we maintain the same four categories used in Paper~I: granules (bright/non-magnetic), magnetic bright points (MBPs), and both magnetic and non-magnetic intergranular lanes. 

The four granulation components are shown in Figure~\ref{fig:2_80_4comp} for centre-to-limb angles: 0$^{\rm{o}}$, 20$^{\rm{o}}$, 40$^{\rm{o}}$, and 60$^{\rm{o}}$; average profiles over the time-series are shown in thick, dark lines and components from individual snapshots are shown in lighter, narrower lines\footnote{The overall shapes of these components (at disc centre) can be compared with Figure~8 in \cite{beeck15b}, where the authors select profiles from four regions of different magnetic field strength and brightness -- however, note these authors simulate a K0V star with average magnetic field of 500~G, and synthesise the Fe~I~$6173~\mathrm{\AA}$ line.}. Note that each individual profile has been shifted in velocity to remove some of the effects from the oscillations, further discussed in Section~\ref{subsec:rmv_osc} (visually this is a small effect, with shifts $< 0.0042~ \AA$); most of the remaining (horizontal and vertical) scatter is due to imperfect removal of the oscillation signal.  

\subsection{Removing the Oscillations}
\label{subsec:rmv_osc}
As mentioned in Section~\ref{subsec:osc_fullprfl}, the oscillations induce variations in both the line profile centres and shapes. At present, we operate under the hypothesis that over the stellar disc these oscillation-induced variations will average out over an appropriately chosen exposure time, but that the granulation variations will still be present. This hypothesis is based on the fact that the granules, as observed on the Sun and in the simulations, tend to appear and disappear in the same locations, making the granulation noise more correlated over time than that from the p-modes. Testing the extent to which this hypothesis stands will be the subject of a future study. As such, herein we seek to remove the impact of the oscillations on our granulation parameterisation; this is especially important given that the oscillation induced variability is orders of magnitude larger than that from the granulation in our simulation box. 

To remove the oscillations, we follow the procedure outlined in Paper~I, where each component profile is shifted by its mean bisector to a rest velocity determined by the respective average bisector position over the time-series. Once the profiles have been shifted, the respective individual components are averaged together to create four granulation component line profiles. Shifting the profiles before creating the final time-average helps prevent the component profiles from being skewed by oscillation-induced broadening (depending on the line depth, the average component profiles were $\sim$1-10~m~s$^{-1}$ wider before removing the oscillation -- with the largest differences at the tops of the profiles). The line shape parameters also vary due to the oscillations; this can been seen in the remaining scatter of the individual profiles about their means in Figure~\ref{fig:2_80_4comp}. However, if the bulk of the movement in the wavelength/velocity domain has been removed then the shape changes should largely average out over the time-series without significant skewing of the profile shapes. 
\begin{figure}[t!]
\centering
\includegraphics[trim=0.cm 0.1cm 0.25cm 0.5cm, clip, width=8.5cm]{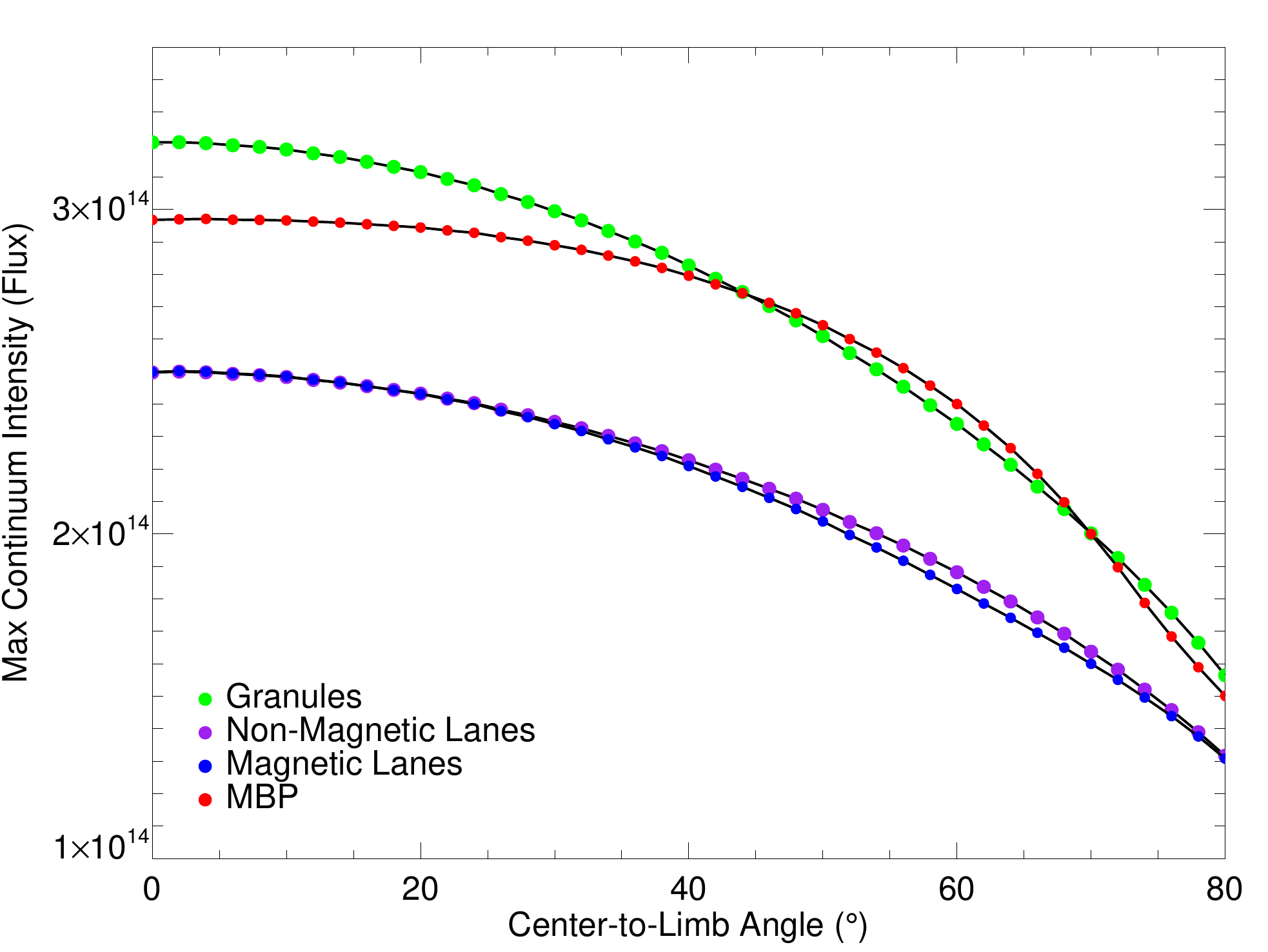}
\caption{The maximum continuum intensity of the four granulation components as a function of limb angle; note the magnetic and non-magentic intergranular lanes values are nearly identical. The initial near constant brightness of the MBP component may be due to inclusion of faculae.} 
\label{fig:2_80_maxcont}
\end{figure}

\subsection{Centre-to-limb Variations}
\label{subsec:2_80_CLV}
Over the stellar disc, all four components experience changes in line profile shape and net velocity. The behaviour between the magnetic and non-magnetic components differ slightly due to the Zeeman splitting in the magnetic components. The non-magnetic components experience a continual decrease in continuum intensity (from limb darkening), shown in Figure~\ref{fig:2_80_maxcont} (and also visible in Figure~\ref{fig:2_80_cont_vel}); these profiles also decrease in line depth/contrast, as well as increase in line width. The changes in line width/depth are due to increasing contributions from plasma flows that were horizontal at disc centre. On the other hand, the magnetic components initially become narrower and deeper once the LOS moves away from disc centre. This is likely because the LOS traverses inclined magnetic flux tubes and therefore does not penetrate as much magnetic flux as when at disc centre; this results in a decrease of the Zeeman splitting in the line profiles \citep{shelyag15}. Moreover, near the limb we view regions higher in the photosphere, and therefore these components may also experience a decrease in their thermal broadening. The MBP components also do not decrease significantly in continuum intensity until closer to the limb (see Figure~\ref{fig:2_80_maxcont}); this is likely because they change from point-like structures within the intergranular lanes at disc centre (that are bright due to enhanced continuum intensity and decreased radiation absorption) to also include bright regions on the granular walls known as faculae at higher inclinations. While the granule itself is non-magnetic, as the granulation snapshot is inclined high magnetic field concentrations in the intergranular lanes decrease the opacity and allow the LOS to reach the granular wall. Such regions then have high brightnesses due to the high temperature of the granule, yet a high magnetic field measurement due to the LOS traversing regions of high magnetic field within the relatively transparent intergranular lane. Although these regions are better known as faculae and do not necessarily have point-like surface areas, since they are both magnetic and bright we include them in the MBP category across the stellar disc. As a result, the decrease in brightness of the MBPs as the simulations are inclined is partially compensated by the appearance of the new MBPs on the granular walls, until $\sim 40^{\rm{o}}$.

\begin{figure}[t!]
\centering
\includegraphics[trim=0.5cm 0.25cm 0.25cm 0.78cm, clip, width=8.5cm]{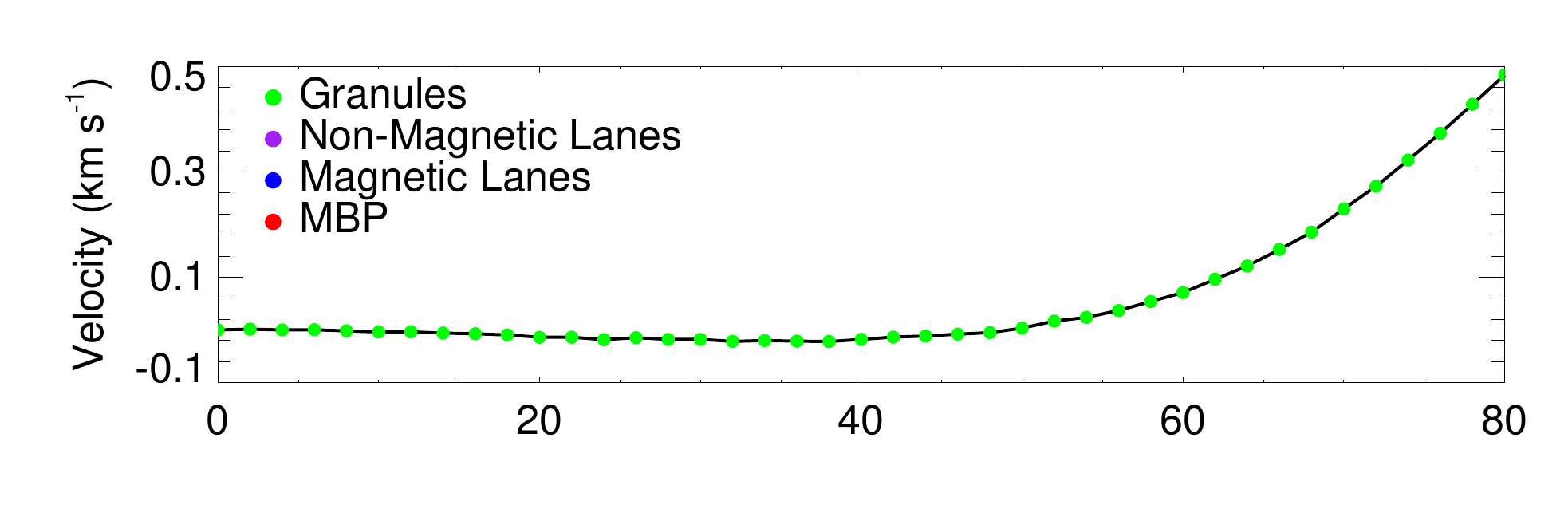}
\includegraphics[trim=0.5cm 0.25cm 0.25cm 0.78cm, clip, width=8.5cm]{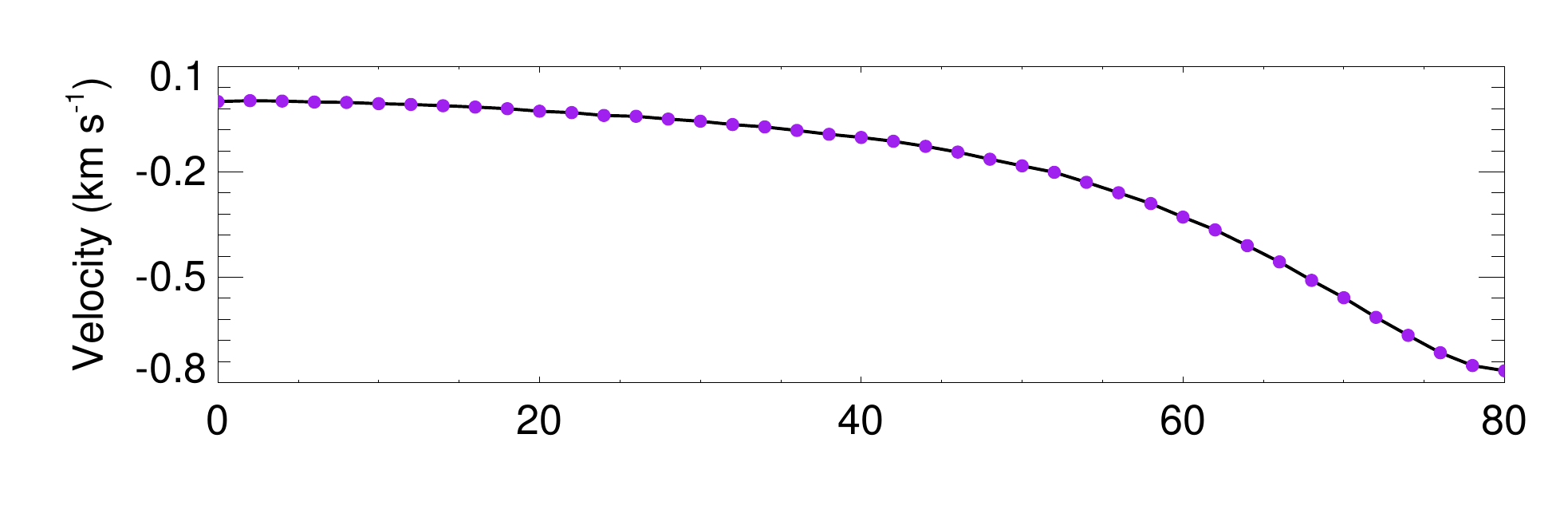}
\includegraphics[trim=0.5cm 0.25cm 0.25cm 0.78cm, clip, width=8.5cm]
{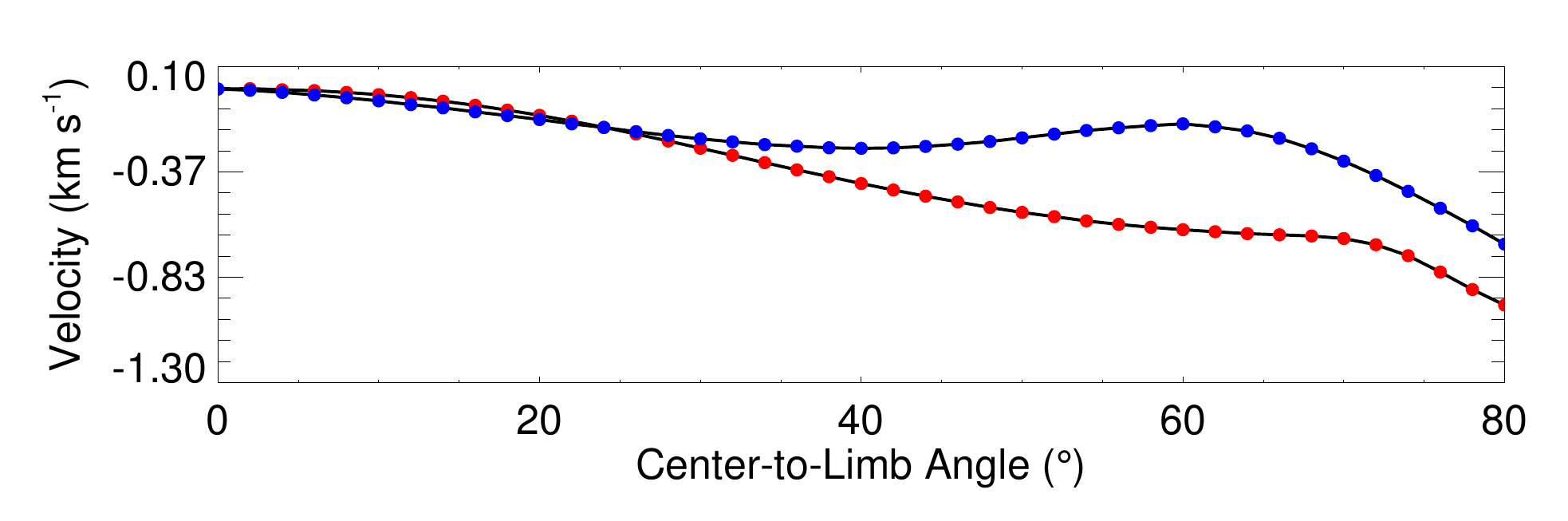}
\caption{The RV of the four granulation components relative to disc centre as a function of limb angle.} 
\label{fig:2_80_rvcomp}
\end{figure}

\begin{center}
\begin{figure}
\centering
\includegraphics[trim=0.5cm 0.25cm 0.25cm 0.78cm, clip, width=8.5cm]{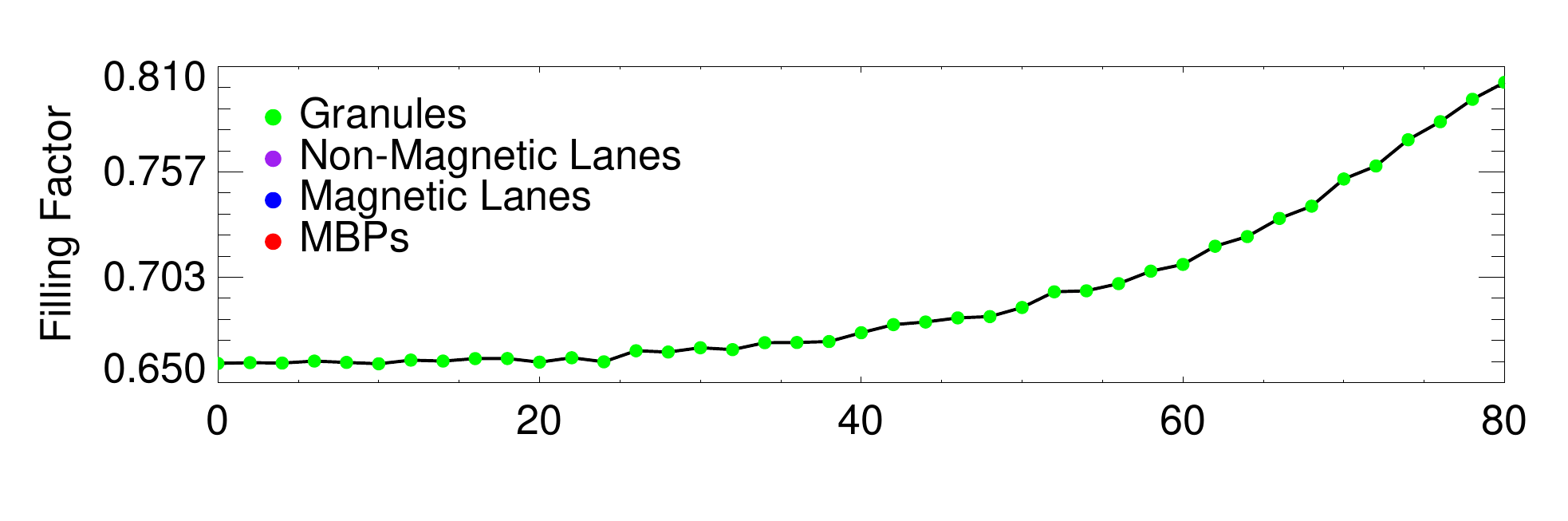}
\includegraphics[trim=0.5cm 0.25cm 0.25cm 0.78cm, clip, width=8.5cm]{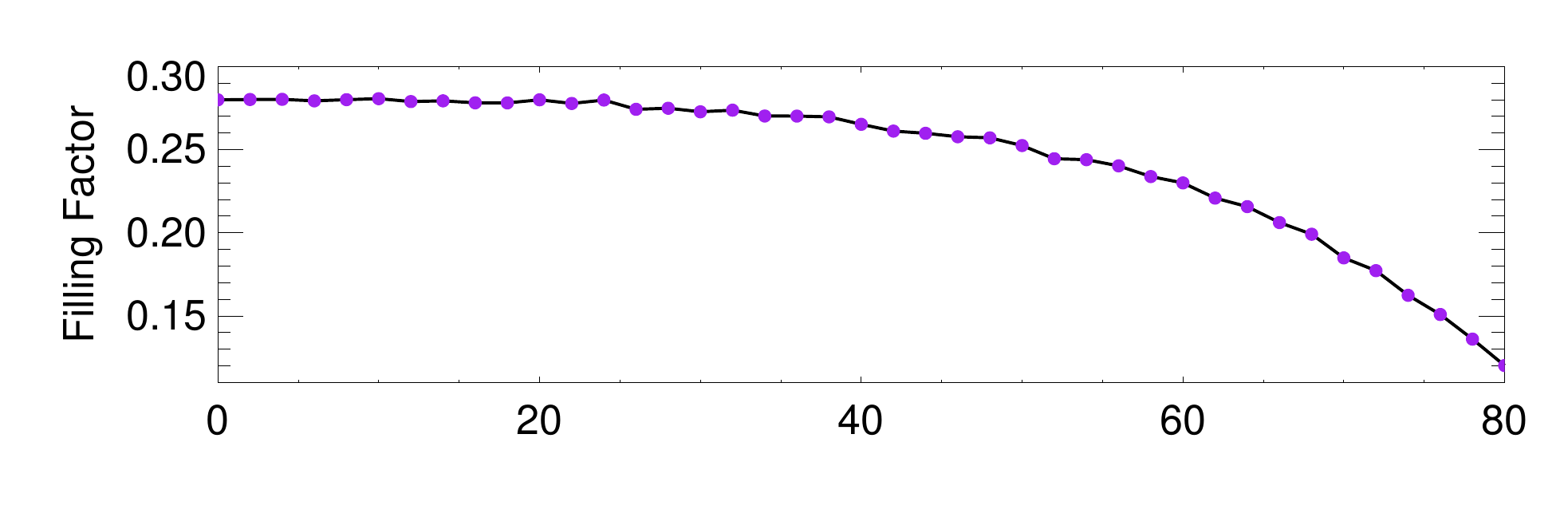}
\includegraphics[trim=0.5cm 0.25cm 0.25cm 0.78cm, clip, width=8.5cm]{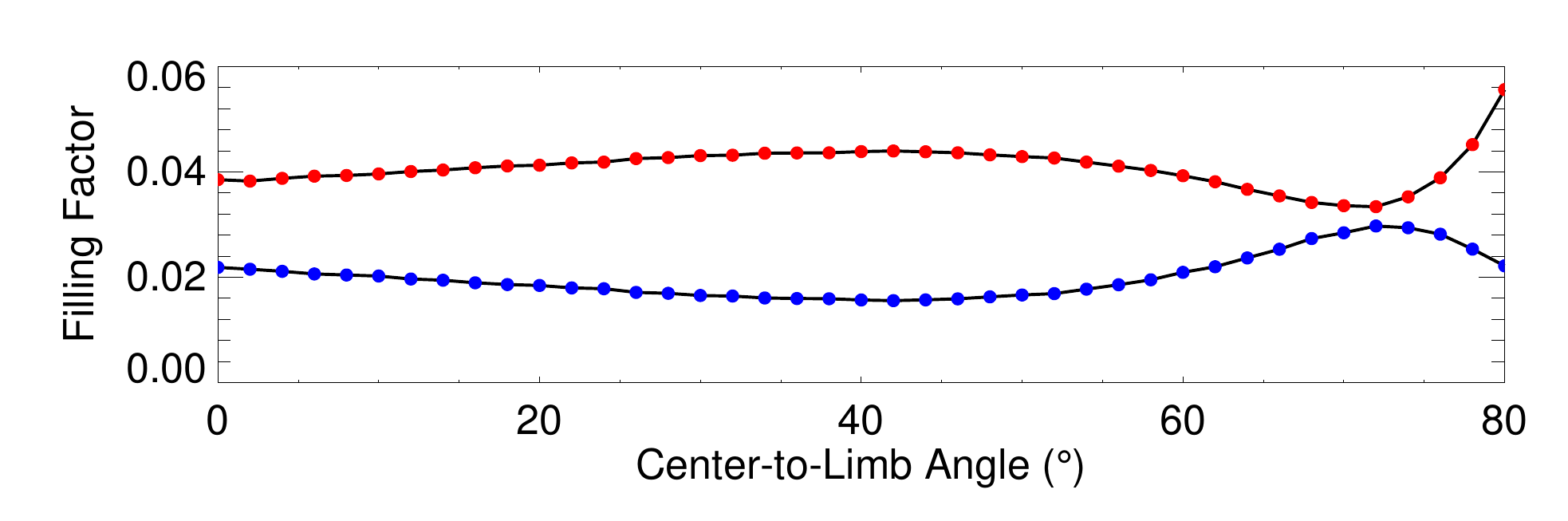}
\caption{Filling factors for the four granulation components as a function of limb angle.} 
\label{fig:2_80_fillfact}
\end{figure}
\end{center}
\vspace{-25pt}

The plasma flows orthogonal to the granule tops also change the net position of the components, resulting in the granule profiles red-shifting and the intergranular lane profiles blue-shifting, relative to disc centre. This is explicitly shown in Figure~\ref{fig:2_80_rvcomp}, where the RVs were calculated in the same manner as Section~\ref{subsec:full_CLV}, but replacing the full line profile in the template with the respective component profiles; it is also visible in the LOS velocity maps shown in bottom of Figure~\ref{fig:2_80_cont_vel}. Some of the vertical LOS velocities at disc centre begin to lie more orthogonal to our LOS as we view granulation towards the limb and thus they decrease in magnitude. In particular, only the near-edge of the granules have high blue-shifts, as seen in Figure~\ref{fig:2_80_cont_vel}, because this is where the plasma has begun to flow into the intergranular lane and now lies more inline with our LOS; the remainder of the top of the granule begins to point away from the observer. The opposite effect is seen in the intergranular lanes, where some of the downward flowing material is now flowing toward the observer as the tiles are inclined; it is also sometimes possible to see the intergranular lanes underneath some of the smaller granules (see Figure~\ref{fig:2_80_cont_vel}). Nonetheless, the granules are still always blue-shifted relative to the intergranular lane components across the stellar disc (see Figure~\ref{fig:2_80_4comp}). 

In addition to the changes in the line shape and position in the four component profiles, there are also variations in their respective filling factors as a function of stellar disc position, shown in Figure~\ref{fig:2_80_fillfact}. From disc centre until $\sim 30^{\rm{o}}$ ($\mu \sim 0.87$), all components remain unchanged within 1\%, indicating the observable field of view does not change significantly. However, from $\sim 40 - 80^{\rm{o}}$ ($\mu \sim 0.67 - 0.17$)), there is a strong decrease in non-magnetic intergranular lane filling factor and subsequent increase in the granule filling factor. We attribute this to increased granular obstruction of intergranular lanes at these angles where the view is dominated by the top regions of the granular `hills' -- an analogy would be standing on top of one hill and looking straight out across several other hills, from this vantage point one can only see the very tops of the hills. Again, we see some evidence for the MBP component capturing faculae as the average filling factor remains roughly constant across most of the disc, until it increases very near the limb. The magnetic intergranular lane filling factor is also roughly constant across the disc; this could be because this component captures regions within the intergranular lane where the magnetic flux is strong, but where the flux tube may not be purely radial. If the magnetic flux tubes are not purely radial, then the magnetic flux present may not be sufficient to completely evacuate the tube and the opacities are not greatly affected; hence these regions are always dark and located within the intergranular lanes (unlike the MBPs). However, the presence of the magnetic flux may be strong enough to alter the opacity in such a way that it counteracts the increased obstruction of the intergranular lanes from the granules. Hence, a constant magnetic intergranular lane component may mean that the visible non-radial flux tubes are roughly constant across the stellar disc. 

\section{Parameterised Reconstruction: Across the Stellar Disc}
\label{sec:2_80_recon}
In order to test the robustness of the parameterisation scheme at different limb angles, we used the four granulation component profiles to reconstruct the original line profiles output from the MHD simulation. We remind the reader that the original profiles were created by averaging together all Stokes~I profiles from each pixel in a given snapshot (i.e. they are the same profiles discussed in Section~\ref{sec:2_80_full_prfls}). Following the methodology outlined in Paper~I, the four component profiles for each limb angle were multiplied by their respective filling factors and summed together to reconstruct the original absorption line profiles output by the simulation. Note that since the parameterisation is created from time-averages, the reconstructed profiles will have no knowledge of the oscillation seen in the original profiles -- only the granulation effects will be captured.

\subsection{Frequency-power Spectra}
\label{subsec:psd}
Since the oscillations alter both the line shape and centre, the original and reconstructed line profiles cannot be compared directly. Moreover, since the hypothesis from Paper~I, on how the oscillations only shift the lines and do not alter their shapes, has been shown to be invalid we should no longer use the average relative error between the original and reconstructed line profiles to evaluate the parameterisation accuracy. As an alternative test, we compare the power spectra densities (PSDs) to the well-known empirical behaviour of solar p-modes and granulation. This is particularly powerful as we can examine the behaviours of both the RVs and the line shape parameters with the same test. 

As mentioned in Section~\ref{subsec:sim}, the cadence of the simulation is close to 30~s, with nearly a third of the snapshots having a cadence closer to $\sim$15~s and each snapshot is taken at instantaneous moments in time (i.e. with zero integration/exposure time). To prevent the non-uniform cadence impacting the PSD, we linearly interpolate the RVs onto a time grid with a fixed cadence of 30\,s and compute frequency-power spectra of these interpolated data; since the granulation and oscillations evolve on timescales of a few minutes, this interpolation should not alter the nature of the data.

Next we fit the power spectra with the same kind of parametric models used for disc-integrated solar (and stellar) observations, using maximum likelihood estimators to optimise the fit. As the reconstructed RVs should contain solely granulation, we fit their power spectrum with a a single ``super-Lorentzian" of the form
 \begin{equation}
 \label{eqn:Pgran}
 P_{\rm gran} = a_0 / (1 + (2\nu / a_1)^{a_2}),
 \end{equation}
where: $\nu$ is frequency; $a_0$ is the maximum power spectral density; $a_1$ calibrates the fall-off of the granulation power with increasing frequency; and $a_2$ is the exponent of the power-law. Then as the original RVs contain both granulation and p-modes, we fit the original RVs with the sum of a Lorentzian (for the p-modes) and the aforementioned super-Lorentzian (for the granulation), i.e.,
 \begin{equation}
 \label{eqn:Ptot}
 P_{\rm tot} = c_0 / (1+(2(\nu - c_2)/c_1)^2) + P_{\rm gran},
 \end{equation}
where: $c_0$ is the maximum power spectral density (or height) of the Lorentzian; $c_1$ is the FWHM of the Lorentzian; and $c_2$ is the central frequency of the Lorentzian.
\begin{center}
\begin{figure}
\centering
\includegraphics[trim=0.05cm 0.1cm 1.9cm 1.4cm, clip, width=8.5cm]{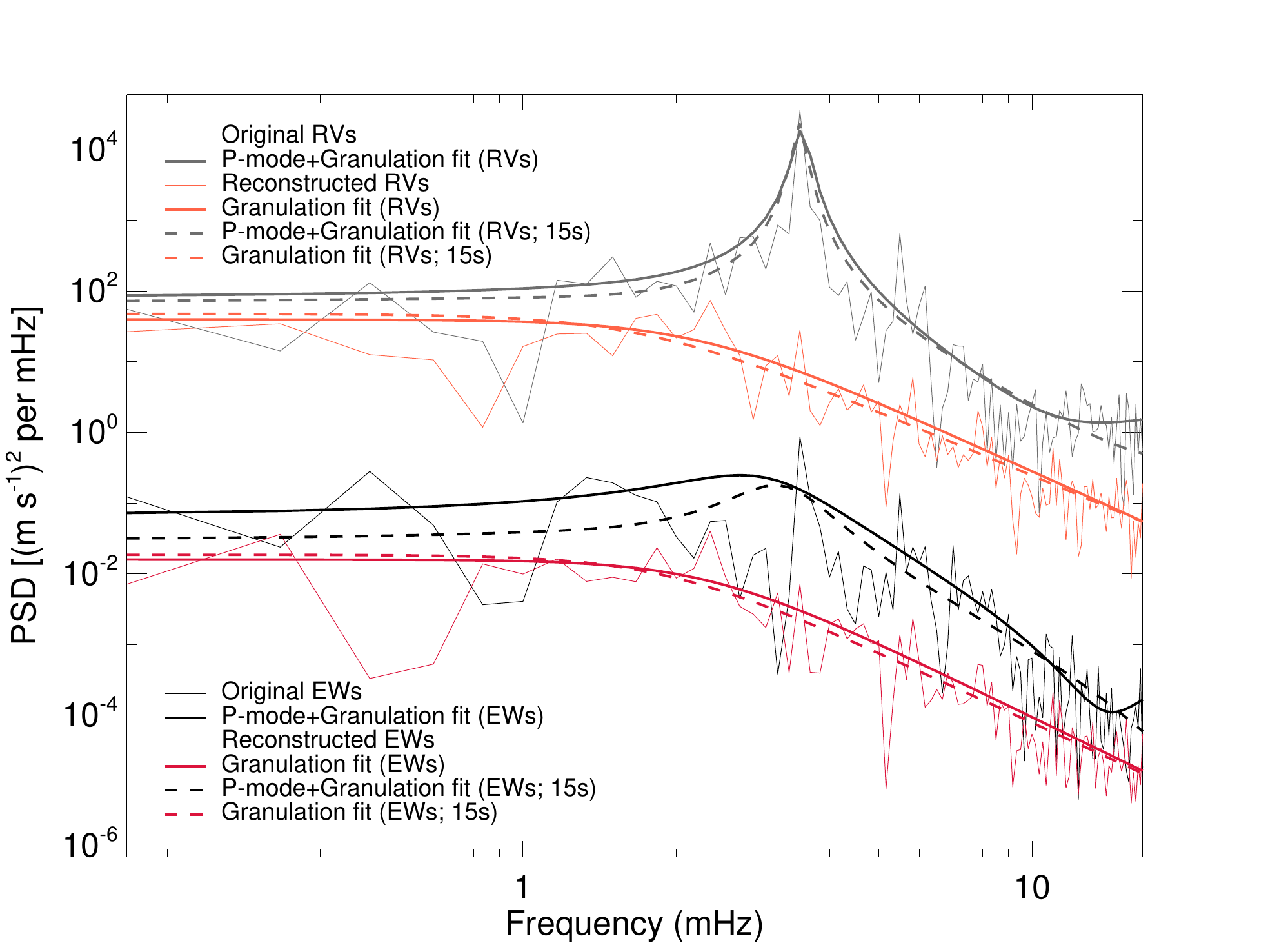}
\caption{The power spectra, at disc centre, of the RVs and equivalent widths from the original and reconstructed line profiles; the equivalent width data has been scaled by a factor of 10$^{-3}$ for viewing ease. Solid lines display an interpolation onto a 30~s cadence and dashed lines for a 15~s cadence; thick lines are the fits to the data (thin lines).} 
\label{fig:PSD_rv_ew}
\end{figure}
\end{center}
\vspace{-20pt}

If the parameterisation sufficiently captures the granulation effects, then we can obtain a good fit to the original time-series with Equation~\ref{eqn:Ptot} whilst holding the super-Lorentzian (Equation~\ref{eqn:Pgran}) fixed to the values obtained from the reconstructed RV fit; as such, we do exactly that to test this hypothesis. 

We perform these fits for data at disc centre, as this is where the influence of the p-modes is most pronounced; hence, an adequate fit here should be sufficient to establish that we can model both the p-modes and the granulation in the power spectrum. Since the original MHD simulation captures just a small region on the stellar disc, we cannot compare directly the best-fitting power parameter ($a_0$) to empirical data. However, we would hope that the characteristic timescale for the granulation ($\tau = 1 / (\pi a_1)$) and the power-law exponent ($a_2$) are both similar to solar observations.

As shown in the top of Figure~\ref{fig:PSD_rv_ew}, we are indeed able to obtain good fits for both the reconstructed/granulation RVs and the original/total RVs. Near very high frequencies, the fit to the total RVs starts to turn over towards higher power; to examine if this behaviour was related to the interpolation scheme, we tested both a spline interpolation (rather than linear) and an interpolation onto a 15~s cadence. The spline interpolation produced very similar results (albeit with a slightly worse $\chi^2$/BIC), but the shorter-cadence results (shown as dashed lines in Figure~\ref{fig:PSD_rv_ew}) show that the fit continues to fall off in power as expected. Hence, the small upturn in the original RVs' fit near high frequencies is most likely due to the limitations of the interpolation and is not indicative of the MHD simulation. The granulation timescale ($\tau = 214 \pm  23$~s) and exponent ($a_2 = 3.3 \pm 0.2$), derived from the fit of Equation~\ref{eqn:Pgran} to the reconstructed data, also agree well with known solar observations (206-215~s and 3.6-4.6, respectively; \citealt{michel09, kallinger14}). We note also that the p-modes in the original simulation box are centred on a similar frequency to the known empirical solar p-modes, $c_2 = 3.5 \pm 0.06$~mHz ($\nu_{max, \odot} = 3.14$~mHz; \citealt{kallinger14}).

In addition to the whole-sale Doppler shifts induced by the p-modes, they also induce changes in the line shape. To examine if our parameterisation captures the correct line shapes, we repeated the above procedure on the equivalent widths (EWs); these are shown alongside the RV results in Figure~\ref{fig:PSD_rv_ew}. The best-fitting timescale and exponent for the reconstructed profile EWs agrees very well with the equivalent fit from the RVs ($\tau = 208 \pm  21$~s, $a_2 = 3.5 \pm 0.2$, respectively). Consequently, we find that the reconstructed profile shapes are following the same behaviour as their RVs, and as such we can assert that we also capture the granulation induced line shape changes with our parameterisation. The fit to the original profile EWs is not quite as good as that for the original RVs; however, this is likely due to the fact that the granulation has a larger impact on the profile shapes than the line centres, as compared to the p-modes, and hence the envelope from the p-modes has less power in the EW PSD and is therefore less well defined. Additionally, a comparison with the granule filling factor PSD shows the same peak frequency at $\sim$2.3~mHz that was found in the reconstructed RVs and EWs PSD; a peak at $\sim$2.3~mHz is also visible in the original PSDs but at a much lower power than the p-mode frequencies. This adds further evidence that the same frequency content from the granulation parameterisation can be seen in the original simulation, but that it is swamped by the large-scale variations induced by the p-modes. 
 
\subsection{Residual RVs}
\label{subsec:2_80_errors}
To further examine the accuracy of the parameterisation in capturing granulation physics, we can still use the hypothesis from Paper~I regarding the residual RVs: if the parameterisation captures the granulation physics, then a cross-correlation between the original average line profiles and their respective reconstructed profiles should provide the RV shifts from the total oscillation signal. Using a cross-correlation technique here allows the oscillation-induced line shape changes to contribute to the oscillation-induced RV shifts. As such, we can still obtain the original granulation RVs from the MHD simulation by subtracting off the oscillation RVs from the total original RVs. See Paper~I for further justifications. Hence, we can examine the residuals between the granulation-induced RVs determined from both the original and reconstructed profiles.

\begin{center}
\begin{figure}
\centering
\includegraphics[trim=0.1cm 0.25cm 0.25cm 0.5cm, clip, width=8.5cm]{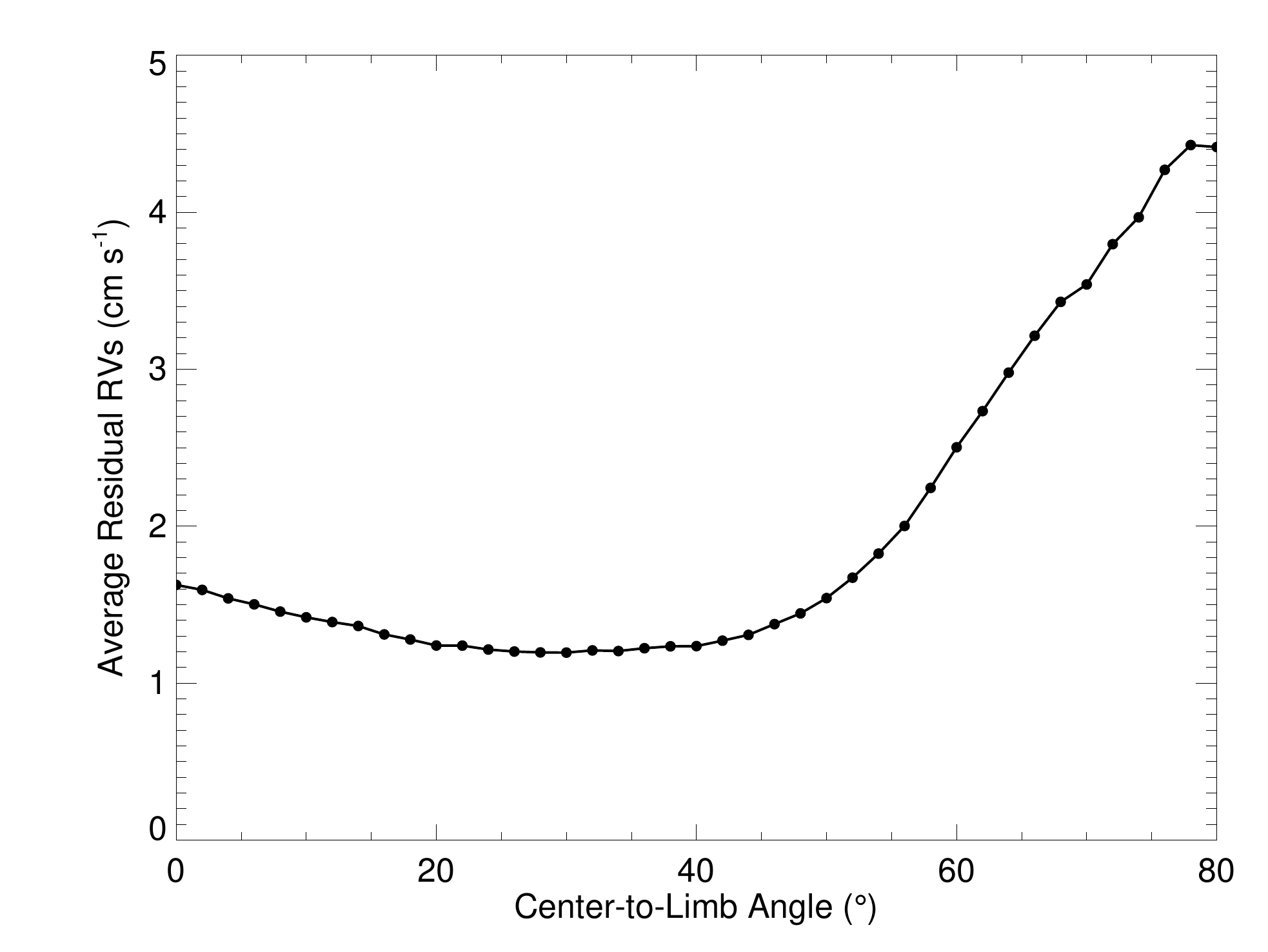}
\caption{The average residual (original - reconstructed) RVs of the time-series as a function of limb angle.} 
\label{fig:2_80_residrv_ang}
\end{figure}
\end{center}

\begin{center}
\begin{figure*}[b!]
\centering
\includegraphics[trim=0.8cm 1.1cm 1.7cm 1.0cm, clip, scale=0.83]{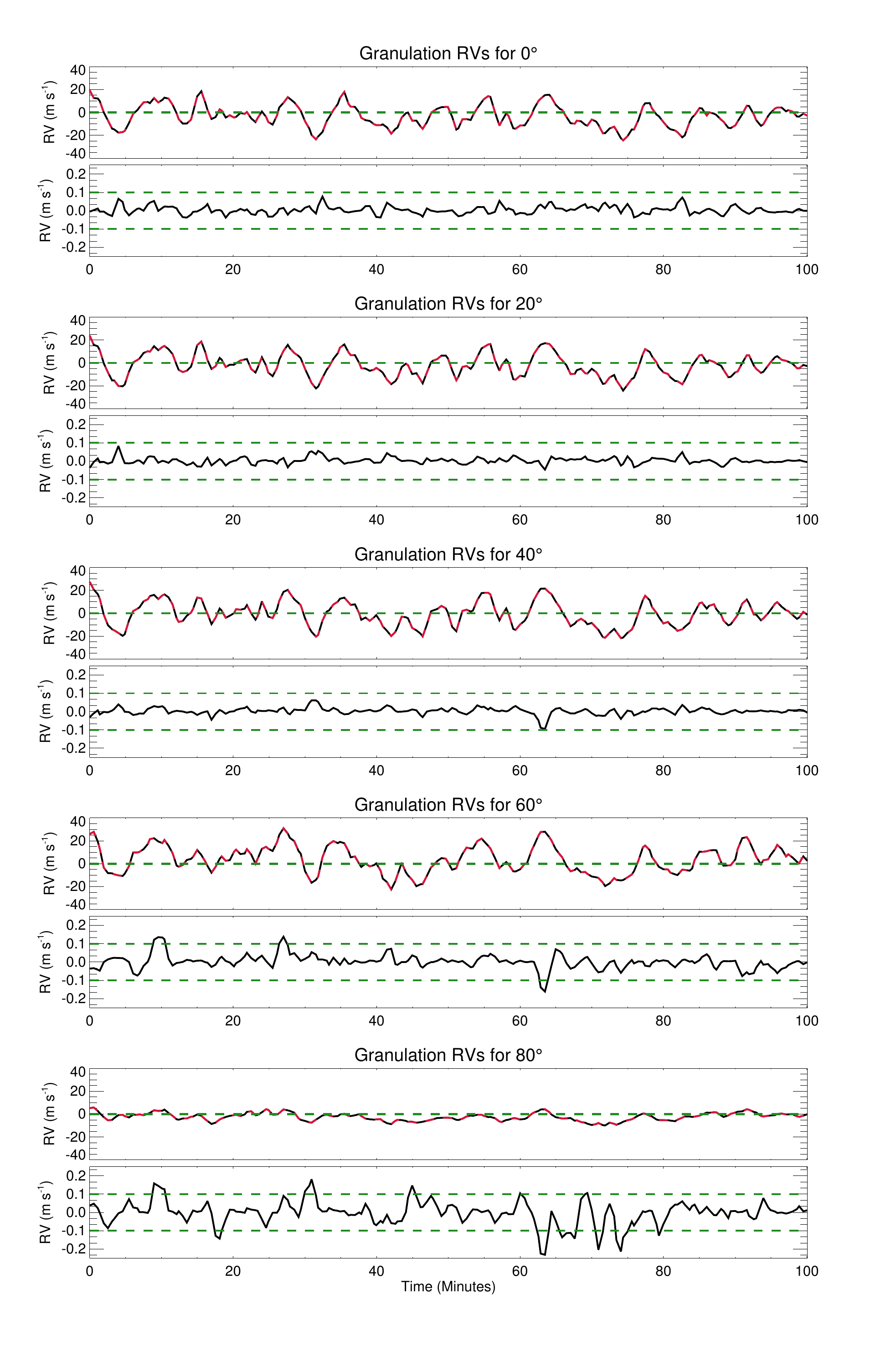}
\caption{The granulation RVs from the simulation (original -oscillation) are shown in black, and those from the reconstruction are shown in dashed red lines, for various centre-to-limb angles; residuals between the two are also shown in black solid lines below. Horizontal dashed lines at $\pm 10 \ \mathrm{cm~s^{-1}}$ are provided to guide the eye.} 
\label{fig:2_80_residrv_time}
\end{figure*}
\end{center}

For this analysis, the RVs originate from a cross-correlation with a template (chosen from the reconstructed time-series) at the same centre-to-limb angle, as we are interested in the RVs at each limb angle and not relative to disc centre; the same template is used with both the original and reconstructed profiles. We found the residuals were always $\le \pm$10 $\mathrm{cm~s^{-1}}$, until very near the stellar limb, but even at 80$^{\rm{o}}$ the residuals are still less than 5 $\mathrm{cm~s^{-1}}$ on average -- see Figure~\ref{fig:2_80_residrv_ang}. The granulation RVs and the residuals between the original and reconstruction for centre-to-limb angles 0$^{\rm{o}}$, 20$^{\rm{o}}$, 40$^{\rm{o}}$, 60$^{\rm{o}}$, and 80$^{\rm{o}}$ are presented in Figure~\ref{fig:2_80_residrv_time}, and shows how well the parameterisation captures the granulation physics at each snapshot.
\vspace{-20pt}

It is possible that some of the residual RVs could arise from errors in the filling factors due to misidentification of components. Since the four average components are composed of averages from numerous profiles within a single snapshot, as well averages throughout the time-series, the average component profiles are likely quite robust, but small changes in the filling factors of a single snapshot could make the reconstruction slightly inconsistent with that particular granulation pattern. If this were the case, this would not be an issue for our parameterisation because it would only mean that in those instances the reconstruction does not match perfectly the original simulated line profile, and not that the technique produces unrealistic granulation line profiles. Additionally, since the residual RVs are very small, and the reconstructions very accurate, the effect of any misidentification of component filling factors in the overall distribution of filling factors from the time-series should be negligible. 
\vspace{-20pt}

\begin{center}
\begin{figure*}[t!]
\centering
\includegraphics[trim=22.9cm 8.7cm 3.5cm 4.5cm, clip, scale=0.45]{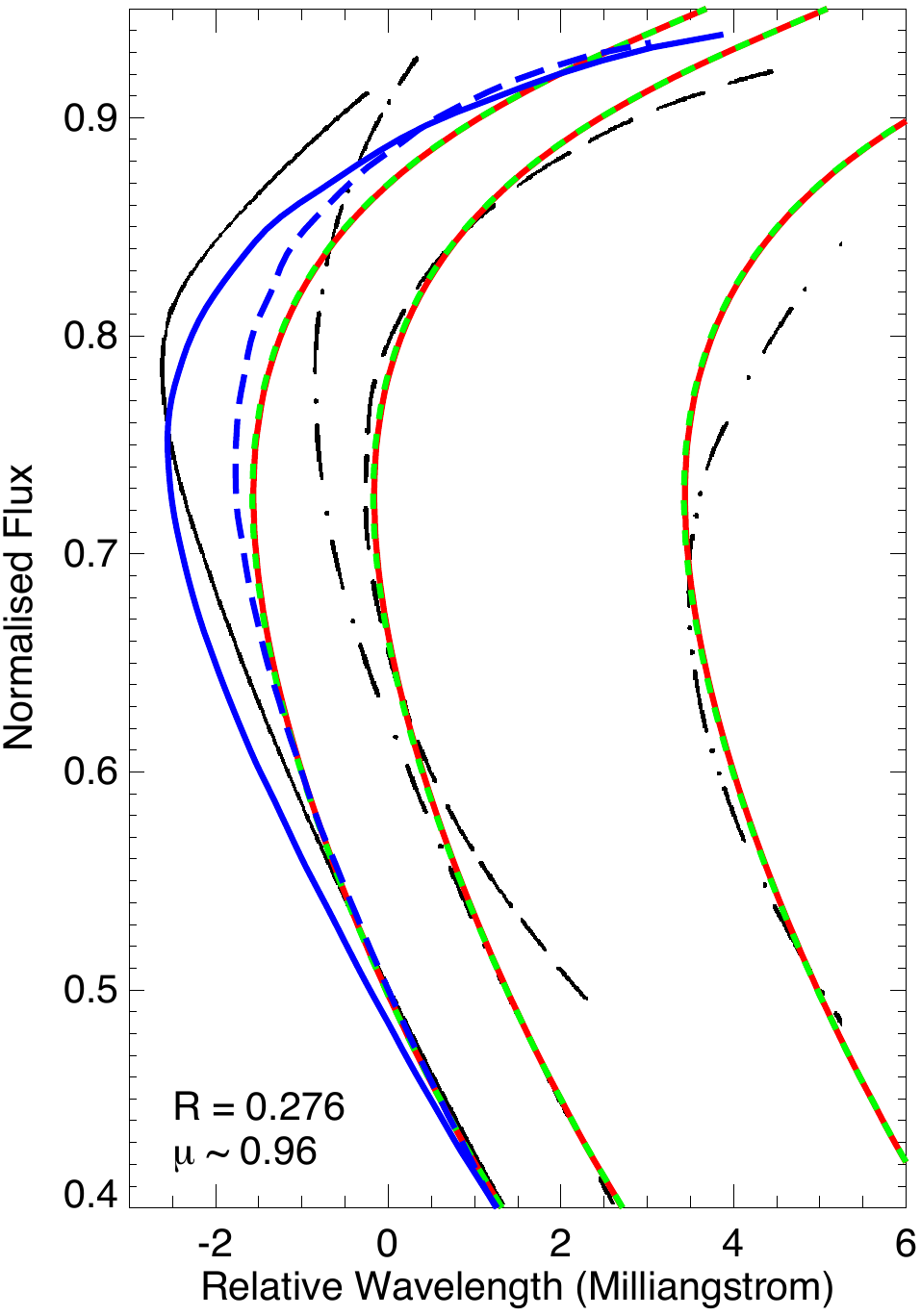}
\includegraphics[trim=22.9cm 8.7cm 3.5cm 4.5cm, clip, scale=0.45]{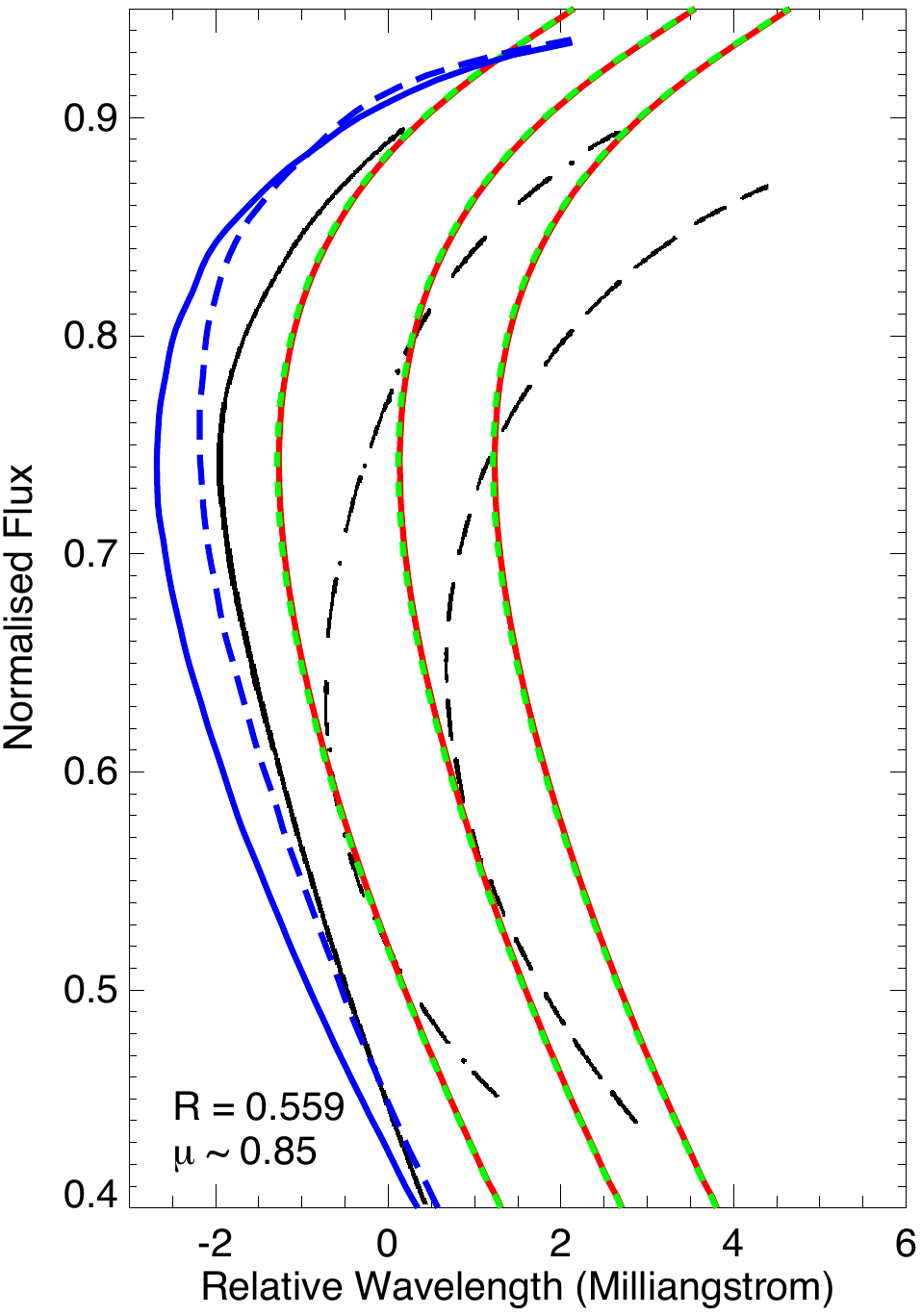}
\includegraphics[trim=22.9cm 8.7cm 3.5cm 4.5cm, clip, scale=0.45]{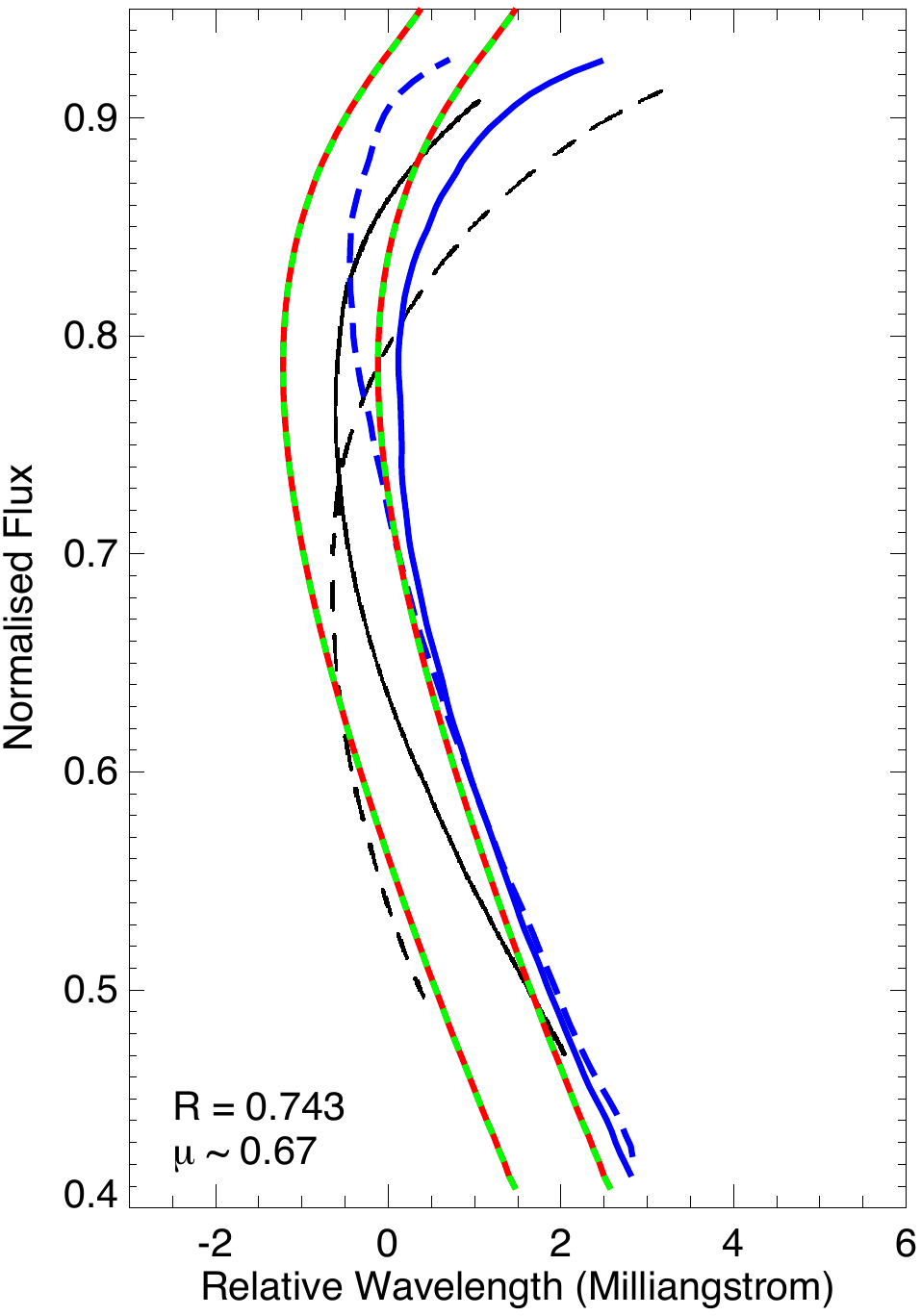}
\includegraphics[trim=22.9cm 8.7cm 3.5cm 4.5cm, clip, scale=0.45]{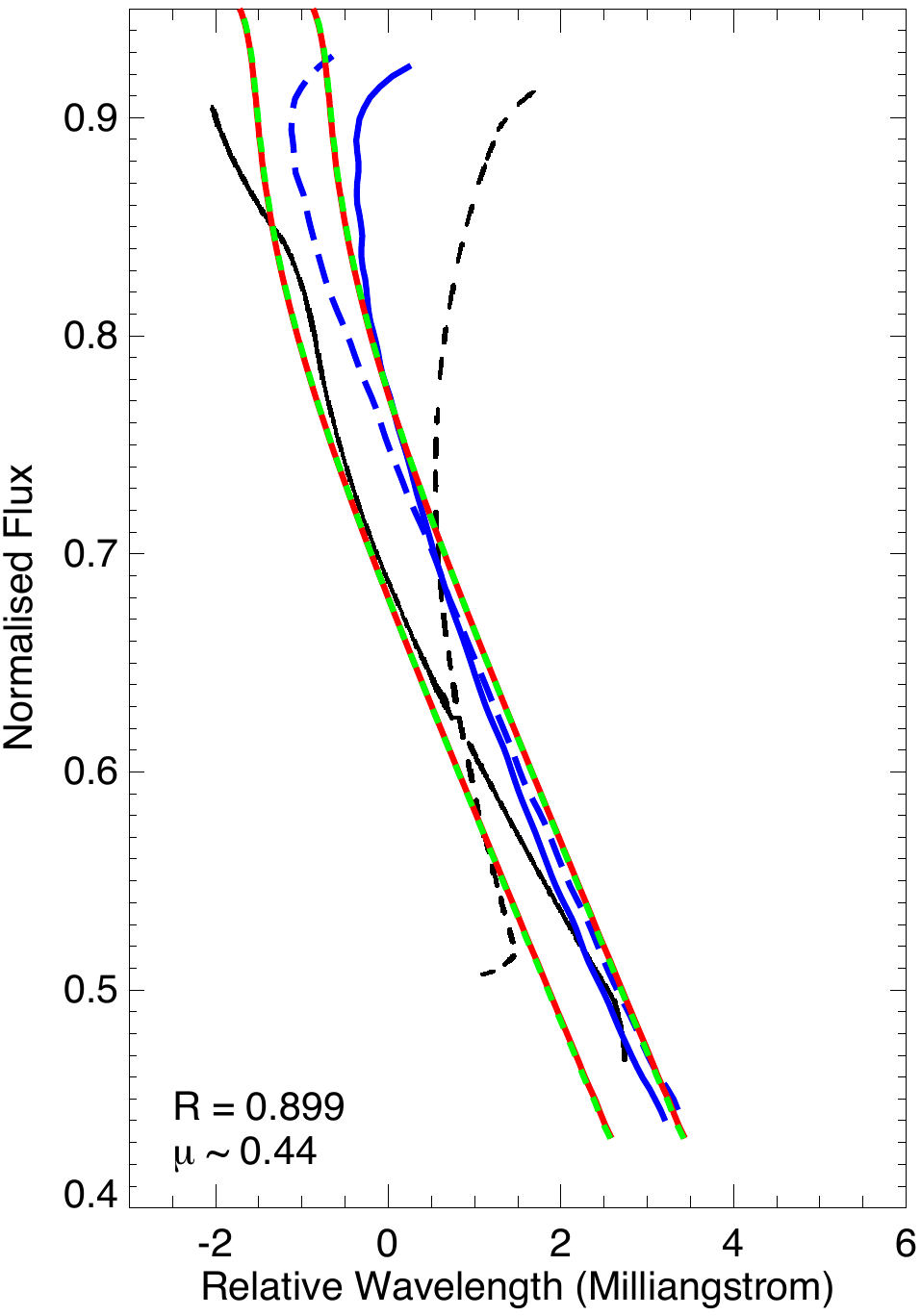}
\caption{Time-average bisectors from the MHD simulation (red are from the original profiles and dashed green lines are from the reconstruction) for various centre-to-limb positions (denoted in units of solar radii and approximate $\mu$), after convolution with an instrumental profile matching \cite{cavallini85a}. Observations from \cite{cavallini85a} are shown in solid black lines for the quiet Sun and dashed/dot-dashed lines for facular regions; observations from \cite{lohner18} for the quiet Sun are shown in blue at $\mu$ intervals of 0.1; dashed lines indicate $\mu$ lower by 0.1. The simulated profiles and data from \cite{lohner18}  have been shifted to various positions to ease shape comparison with \cite{cavallini85a}. 
} 
\label{fig:bi_cav}
\end{figure*}
\end{center}
\vspace{-30pt}

It is also possible that some of these residual RVs could originate not from fault in the logic of the parameterisation nor lack of optimal simulation resolution or length, but from random small-scale vortices that originate in the simulations and are averaged out in the creation of the four granulation component profiles. Random small-scale vortices are a by-product of photospheric magneto-convection and appear in the radiative 3D MHD solar simulations. These small-scale vortices are short-lived and unpredictable. The presence of any such vortices at a particular point in time will introduce small line asymmetries in the average line profile of the corresponding snapshot in the simulations. Since our four component profiles are created by averaging components in a given category over the time-series, it is impossible for these vortices not to be averaged out. However, this is not a problem for our parameterisation because its purpose is to be used to simulate model Sun-as-a-star observations of stellar surface granulation, and such disc-integrated observations would also average out the RV variations from these random, short-lived, small-scale vortices \citep{shelyag11}. 

\section{Comparison to Observations}
\label{subsec:obs}
Here we compare our simulated line profiles with solar observations; for a comparison between different HD/MHD codes we direct the readers to \cite{beeck12}\footnote{Note, Figures~\ref{fig:clv_core} and \ref{fig:clv_hmi} can be tentatively compared to Figure~18 in \cite{beeck13} and Figure~11 in \cite{beeck15b}, but note the differences in magnetic field, line choice, and RV calculation.}. Unfortunately, there are no known observations of the Fe~I~$6302~\mathrm{\AA}$ line that provide both line profile shape and net convective blueshift measurements where the magnetic field is known to be similar to our simulation. As such, we expect some differences when comparing to the literature due to variations in magnetic field strength as compared to the MHD simulation.

\subsection{Centre-to-limb Line Bisectors}
\label{subsec:obs_bi}
Foremost, we compare line profile shapes by inspecting the line bisectors at various limb angles and comparing them to the solar observations from \cite{cavallini85a} and \cite{lohner18}. Since both of these authors use time-averages (either via binned data or long exposures) to smooth out the impact of the p-modes (and increase signal-to-noise), we make this comparison on a line profile created from averaging over the entire simulation time-series. Moreover, as we are comparing purely the line shape in this instance, we shift the simulated profiles in wavelength to closer examine the bisectors variations (see Figures~\ref{fig:clv_mn_bi} and \ref{fig:clv_core} for a comparison of net RV). Finally, we convolve our simulated profile with an instrumental profile matching the observations of \cite{cavallini85a} (a Gaussian with a full-width at half-maximum [FWHM] of $40~m\mathrm{\AA}$); this is particularly important as the symmetric instrumental profile acts to smooth out some of the asymmetries in the underlying Fe~I~$6302~\mathrm{\AA}$ profile. In Figure~\ref{fig:bi_cav}, we can see that our simulated profile shares characteristics of both the observed quiet Sun and facular region profiles. Near disc centre, our profiles and those of the observed quiet Sun share a similar blueward slope near the bottom of the line; at the same position, our simulated profiles are more similar to observed facular region profiles in the redward slope seen in the upper part of the line. Towards the stellar limb, the simulated profile begins to share even more curvature similarities with the quiet Sun observations as compared to the facular regions. This behaviour likely indicates that the magnetic field in our simulations is greater than the quiet Sun (as expected), yet smaller than the facular regions observed by \cite{cavallini85a}. Nonetheless, given the known differences in magnetic field strength, we find good line shape agreement with the observations across the stellar disc. 

\subsection{Centre-to-limb Convective Blueshift}
\label{subsec:obs_cb}
Next we compare the CLV in the net convective blueshift as a function of limb angle. For the Fe~I~$6302~\mathrm{\AA}$ line, we have CLV measurements for the quiet Sun from \cite{cavallini85b}, and even higher precision measurements from \cite{lohner18}. However, as we know the enhanced magnetic field in our simulation inhibits the convective flows, we anticipate potentially large differences in the CLV. 

As such, we have also synthesised a small subsample of Fe~I~$6302~\mathrm{\AA}$ line profiles for a MHD simulation with a net magnetic field of 0~G. This sample includes profiles from six snapshots, separated in time by 20 minutes each, at nine centre-to-limb positions, in steps of $\sim 0.1 \mu$ (from 1.0 - 0.1). In this way, this subsample should be sufficient to average out much of the impact from the stellar oscillations, as well as capture the overall centre-to-limb behaviour. A complete characterisation of this simulation is beyond the scope of this paper and will be the subject of future work; its purpose here is simply to investigate the CLV in net convective blueshift as compared to the 200~G simulation presented throughout this work and the empirical quiet Sun observations.

On the other hand, to compare the 200~G simulation with solar observations at different magnetic field strengths, we turn to observations of the Fe~I~$6173~\mathrm{\AA}$ line. Following \cite{haywood16}, \cite{palumbo18} were able to use AIA data to isolate quiet Sun, faculae network, and plage regions, with increased average magnetic field respectively. Then the HMI data was used to extract the local RVs for each region separately. Note these RVs are expressed relative to the quiet Sun in order to overcome various instrumental offsets (such as filters drifting over the Fe line); hence, we can only compare the centre-to-limb shapes and relative shifts between regions of different magnetic field strength. The Fe~I~$6173~\mathrm{\AA}$ line is quite similar to the Fe~I~$6302~\mathrm{\AA}$ line in terms of physical properties (e.g. line strength, excitation potential and line broadening parameters), and nearly identical in magnetic sensitivity (i.e. Land\'{e} factor), and therefore should provide a reasonable comparison; note the Fe~I~$6302~\mathrm{\AA}$ line is slightly stronger, with a slightly larger excitation potential (0.58 vs 0.55 and 3.7 vs 2.2 eV, respectively; \citealt[][and references therein]{ryabchikova15})\footnote{These values were derived from the VALD database assuming solar properties (http://vald.astro.uu.se).}. 

\begin{center}
\begin{figure}[t!]
\centering
\includegraphics[trim=0.5cm 0.25cm 0.25cm 0.5cm, clip, width=8.5cm]{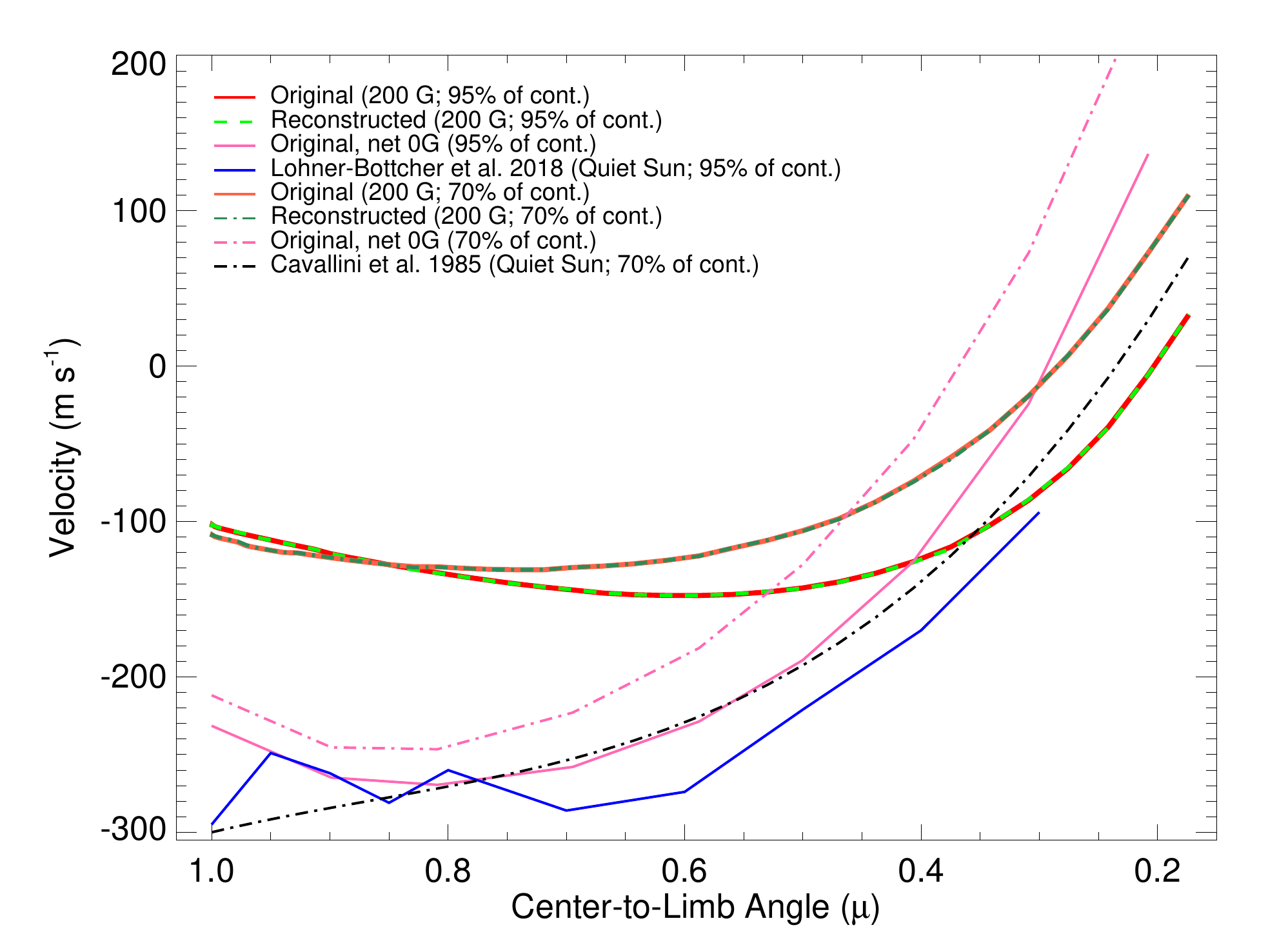}
\caption{The net RV as a function of limb angle, calculated via the mean bisector up to either 70\% or 95\% of the continuum, and convolved with the corresponding instrumental profile; thick lines indicate the 200~G simulation.} 
\label{fig:clv_mn_bi}
\end{figure}
\end{center}
\begin{center}
\begin{figure}[t!]
\centering
\includegraphics[trim=0.5cm 0.25cm 0.25cm 0.5cm, clip, width=8.5cm]{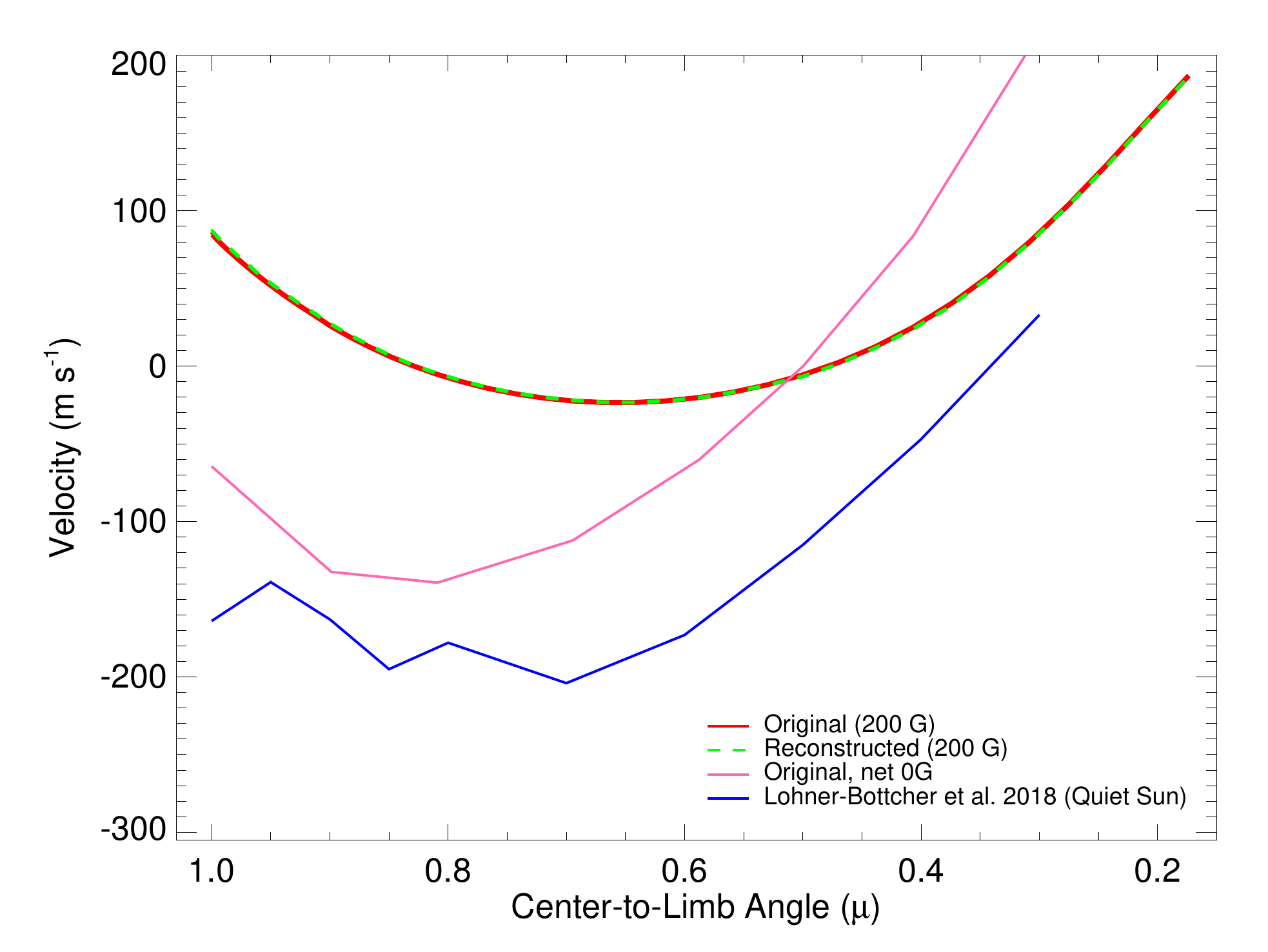}
\caption{The net RV as a function of limb angle, calculated via a parabolic fit to the bottom of the line profile core. Profiles from the simulations were convolved with the instrumental profile from the LARS spectrograph (thick lines indicate the 200~G simulation).} 
\label{fig:clv_core}
\end{figure}
\end{center}
\begin{center}
\begin{figure}[t!]
\centering
\includegraphics[trim=0.5cm 0.25cm 0.25cm 0.5cm, clip, width=8.5cm]{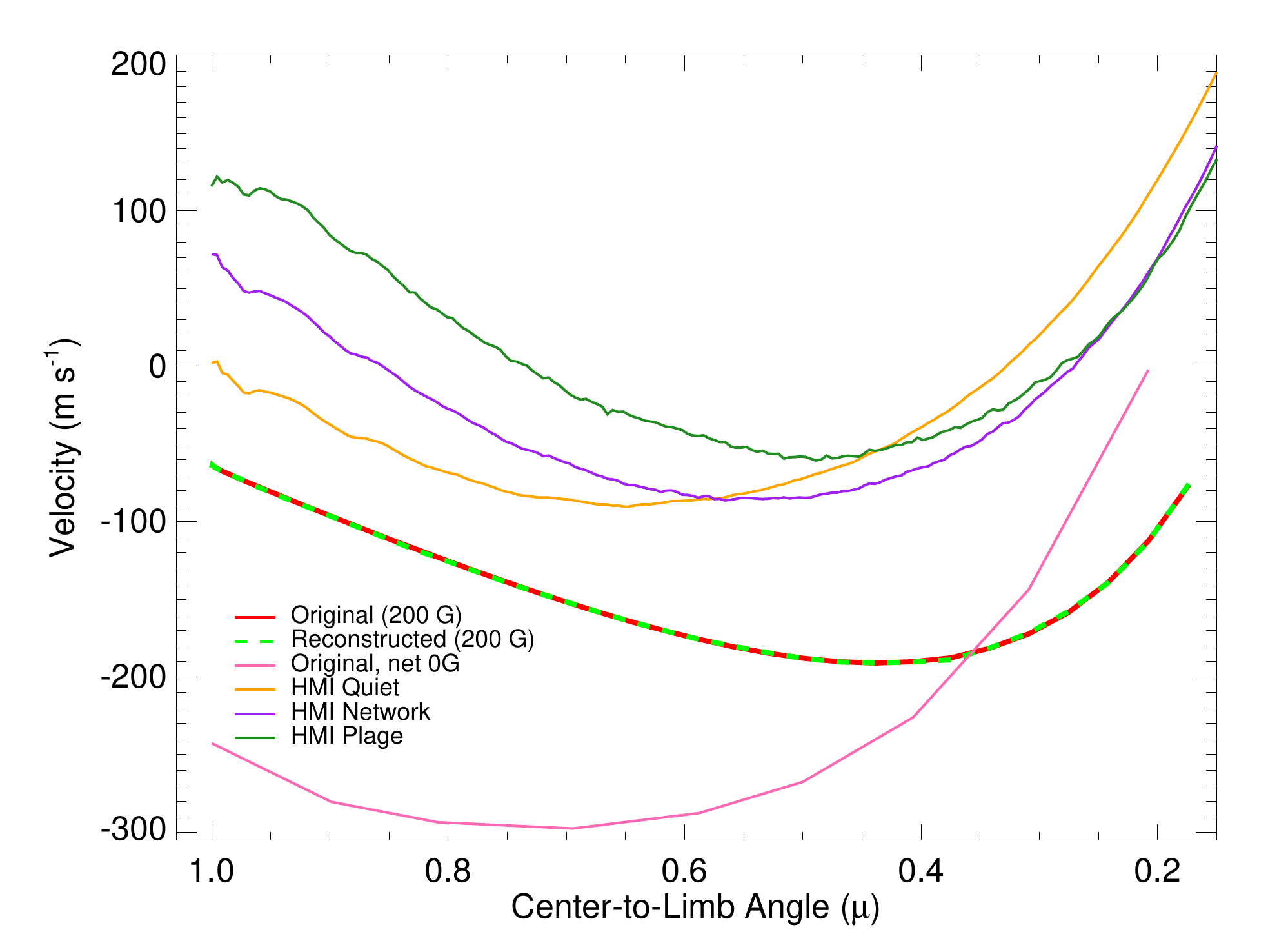}
\caption{The net RV as a function of limb angle, calculated via the first moment. Profiles from the simulation were convolved with the instrumental profile of HMI (thick lines indicate the 200~G simulation). The Fe~I~$6173~\mathrm{\AA}$ HMI observations are separated into contributions from the quiet photosphere, facular network, and plage regions (with increasing magnetic field respectively) and are relative to the quiet Sun.} 
\label{fig:clv_hmi}
\end{figure}
\end{center}
\vspace{-70pt}

The calculation of the net convective shift is particularly sensitive to how the RV is measured, as well as the influence of the instrument profiles. The \cite{cavallini85b} measurements were determined via the mean of the line bisector up to 70\% of the continuum; these measurements were also reported relative to disc centre and therefore do not provide an absolute RV scale. Since the \cite{cavallini85b} results were reported relative to disc centre, we subtracted an offset at 300~m~s$^{-1}$, the expected approximate net convective blueshift of the Sun at disc centre, when making our comparison. \cite{lohner18} provided measurements calculated both via a mean bisector (this time up to 95\% of the continuum) and via a parabolic fit to the line core. Finally, the HMI measurements are reported via a calculation of the first moment. Comparisons to these three techniques are shown in Figures~\ref{fig:clv_mn_bi}, \ref{fig:clv_core}, and \ref{fig:clv_hmi}. For each, we again used the result from a time-average over the entire simulation duration\footnote{For the net 0 G simulation, we averaged over the aforementioned six snapshots, separated in time by 20 minutes each.}, and convolved with the appropriate instrumental profile (a Gaussian with a FWHM of 9, 40, and $76~m\mathrm{\AA}$ for comparisons with \citealt{lohner18, cavallini85b, palumbo18}, respectively. Note, we anticipate the RV determined from the line core to be most sensitive to any residual oscillations that may not be completely averaged out in our limited sample of the net 0~G simulation; this is because it is determined from only a small section of the overall line profile (as compared to the mean bisector or first moment). Moreover, since the oscillations become less coherent in phase near the limb (see Figure~\ref{fig:orgRV}), it may be more difficult to average out the oscillations from randomly selecting snapshots (despite the lower amplitudes  near the limb); hence the net 0~G simulation may differ from the observations more near the limb. Finally, since the magnetic field dampens the convective motions, it naturally also dampens the induced oscillations; as such the lower magnetic field strength simulation may require further averaging to bin out the oscillation impact.

It is clear from Figures~\ref{fig:clv_mn_bi} and \ref{fig:clv_core} that the results from our 200~G MHD simulation are far more redshifted than the quiet Sun observations. The minimum CLV RV, and subsequent steep increase in redshift, also happens closer to the limb in the MHD simulations as compared to the quiet Sun. Both aspects are to be expected since the simulation has a magnetic field strength closer to a facular/plage region than the quiet photosphere, and the enhanced magnetic field retards the plasma flows. This is evidenced by comparisons with the subsample from the net 0~G MHD simulation, and the HMI observations of the quiet Sun, network, and plage regions. In Figures~\ref{fig:clv_mn_bi} - \ref{fig:clv_hmi}, we see in each instance the net 0~G simulation matches much more closely the empirical observations from \cite{cavallini85b}, \cite{lohner18}, and \cite{palumbo18}; where the differences can be attributed to small differences in net magnetic field strength and/or the impact of non-perfect smoothing over the stellar oscillations (see Figure~\ref{fig:orgRV} for the impact of the oscillations on the centre-to-limb behaviour). In particular, we highlight that the larger differences between the 0~G simulation and observations seen in Figure~\ref{fig:clv_core} are in line with what we expect due to contamination from residual oscillation signatures.

Moreover, from the HMI observations in Figure~\ref{fig:clv_hmi}, we can observe the impact of increasing magnetic field by comparing the quiet photosphere to the more magnetic facular network regions to the even more magnetic plage regions. In doing so, we see an increase in magnetic field increases the net redshift, as well as the initial gradient in the CLV; it also pushes the minimum CLV towards the solar limb. This same general behaviour from the HMI observations can be seen when comparing the subsample of net 0~G simulation to the full 200~G simulation. There is some variance, between simulation and observation, in the relative shift between the `quiet' and more `active' regions, but we attribute this largely to differences in the net magnetic field strength and configuration. We note that the main distinction between the simulated $6302~\mathrm{\AA}$ line and observed $6173~\mathrm{\AA}$ line behaviour is the gradient in the initial blueshift; however, this could be due to slight variations in the formation heights of the two lines. Overall, the magnitude and behaviour of the CLV in both line profile shape and net convective blueshift match well the observations, given the known differences in magnetic field strength (and also line properties in regards to HMI). 

\section{Concluding Remarks}
\label{sec:conc}
Throughout this paper we have extended the multi-component parameterisation of magneto-convection from disc centre (Paper~I) towards the stellar limb. Away from disc centre, contributions from plasma flows orthogonal to the granule tops become significant and the geometry of the granular hills and intergranular lane valleys play a vital role in the resultant line profile asymmetries/shifts. In particular, we have examined the CLV in net RV, and broken down the increase in redshift toward the limb into the individual granulation component contributions. We found that the granule and non-magnetic intergranular lane line profiles decreased in line depth and increased in line width towards the limb; while, the magnetic components initially increased in line depth and decreased in line width as they were viewed toward the limb, and subsequently through areas of lower magnetic flux. Near the limb, granules continued to redshift and the intergranular lane components continued to blueshift relative to disc centre -- though the granules were still always blueshifted relative to the intergranular lane components. Moreover, the granule filling factor continued to increase toward the limb as the granules obstructed the intergranular lane components. The MBP component likely captures the hot granular walls (faculae) off disc centre, which prevents its continuum intensity from falling off until closer to $\sim 50^{\rm{o}}$. The result of all this behaviour is due to the relationships between the visible plasma flows and the corrugated nature of the granulation. 

We also compared the results from our simulation to various empirical observations of the Sun. Given the differences in magnetic field strength, we found good centre-to-limb agreement in the bisector shape, in the net RV, the p-mode amplitude/power, and behaviour of the PSD for both the RVs and equivalent width. It was crucial to examine the line shape behaviour, as we found that both the p-modes and the granulation change the line shape. We also determined that our parameterisation can recreate the granulation RV shifts found in the MHD simulation to better than 10~cm~s$^{-1}$ across the stellar disc. Hence, we are confident that our granulation parameterisation captures both the line shape and RV shifts indicative of solar surface convective flows. As such, in a forthcoming paper we will use this parameterisation, and the probability distributions of the filling factors from the time-series herein, to create new granulation profiles representative of those that would be produced through the more computationally intensive radiative 3D MHD simulations. Then these new granulation line profiles will be used to create Sun-as-a-star model observations to examine the granulation induced line shape changes and their relationship with the induced RV shifts. Later we can extend this work to different magnetic field strengths and line profiles with different physical properties. Ultimately, we aim to determine the optimum techniques to remove granulation (and potentially oscillation) induced line asymmetries and RV shifts from exoplanet observations. This work is vital to overcome the barriers of astrophysical noise and detect true Earth analogues in the future. 

\acknowledgments
The authors would like to thank S. Sulis for useful discussions regarding coloured noise and periodograms, and J. L\"{o}hner-B\"{o}ttcher for sharing their results from the LARS spectrograph. The authors also thank the anonymous referee for their useful comments, which improved the clarity of this manuscript. H.M.C. acknowledges the financial support of the National Centre for Competence in Research “PlanetS” supported by the Swiss National Science Foundation (SNSF); S.S., H.M.C and C.A.W acknowledge support from the Leverhulme Trust. W.J.C., G.R.D. and C.A.W. acknowledge support from the UK Science and Technology Facilities Council (STFC), including grants ST/I001123/1 and ST/P000312/1. Funding for the Stellar Astrophysics Centre is provided by The Danish National Research Foundation (Grant agreement no.: DNRF106). S.H.S. was supported by NASA Heliophysics LWS grant NNX16AB79G, and M.L.P. was supported by the NSF-REU solar physics program at SAO, grant number AGS-1560313. Some of this work was performed under contract with the California Institute of Technology (Caltech)/Jet Propulsion Laboratory (JPL) funded by NASA through the Sagan Fellowship Program executed by the NASA Exoplanet Science Institute (R.D.H.). This work used the DiRAC Data Centric system at Durham University, operated by the Institute for Computational Cosmology on behalf of the STFC DiRAC HPC Facility; DiRAC is part of the National E-Infrastructure. This research has made use of NASA's Astrophysics Data System Bibliographic Services, as well as the VALD database operated at Uppsala University, the Institute of Astronomy RAS in Moscow, and the University of Vienna.

\software{MURaM \citep{vogler05,shelyag14, shelyag15};  OPAL \citep{rogersOPAL, rogersnayfonovOPAL}; NICOLE \citep{NICOLE1, NICOLE2}}

\bibliographystyle{aasjournal}
\bibliography{abbrev,mybib}

\end{document}